\documentclass[showpacs, pra, twocolumn,preprintnumbers ,amsmath, amssymb, floatfix, superscriptaddress, aps]{revtex4}
\usepackage{color}
\usepackage{amsmath,amssymb}
\usepackage{pifont}
\usepackage{amssymb}  
\usepackage{bbold}
\usepackage{float}
\usepackage{subfloat}
\usepackage[caption=false]{subfig}
\usepackage{tikz}
\usepackage{makecell}
\usepackage{subfig}
\usepackage{pifont}   
\usepackage{graphicx} 
\graphicspath{{Figures3/}}
\usepackage{dcolumn}  
\usepackage{bm}       
\usepackage{multirow} 
\usepackage{placeins}
\usepackage[colorlinks]{hyperref}
\usepackage{mathtools}

\newcommand{\ket}[1]{\left|{#1}\right\rangle}
\newcommand{\bra}[1]{\left\langle{#1}\right|}

\newcommand{\eqfitpage}[1]{\resizebox{\linewidth}{!}{$#1$}}

\begin{document}
	
\title{Effect of magnetic field and light on energy levels of (1+3+1) chirally twisted multilayer graphene system}
\date{\today}
\author{Nadia Benlakhouy}
\email{nadia.benlakhouy@um6p.ma}
\affiliation{School of Applied and Engineering Physics,  University Mohammed VI Polytechnic, Ben Guerir, 43150, Morocco}
\author{Ahmed Jellal}
\email{a.jellal@ucd.ac.ma}
\affiliation{Laboratory of Theoretical Physics, Faculty of Sciences, Choua\"ib Doukkali University, PO Box 20, 24000 El Jadida, Morocco}
\author{Hocine Bahlouli}
\affiliation{Physics Department and IRC Advanced Materials, King Fahd University of Petroleum and Minerals, Dhahran 31261, Saudi Arabia}
	\author{Pablo Díaz}
               \affiliation{Departamento de Ciencias F\'{i}sicas, Universidad de La Frontera, Casilla 54-D, Temuco 4811230, Chile}  
            		\author{David Laroze}
            \affiliation{Instituto de Alta Investigación, Universidad de Tarapacá, Casilla 7D, Arica, Chile}

\begin{abstract}
We study the Hofstadter butterfly spectrum in (1+3+1) chirally twisted multilayer graphene (CTMLG) subject to perpendicular magnetic field and light with different polarizations. We focus on the interplay between twist angles and light-induced effects. In equilibrium, we examine symmetric ($\theta_1 = \theta_2$) and asymmetric ($\theta_1 \neq \theta_2$) configurations. Our results show that asymmetric configurations cause distinct effects in the electronic energy spectrum. However, the unique symmetry of the system ensures that the spectra remain identical when the twist angles are interchanged. 
This highlights the role of interlayer coupling in shaping the electronic structure of CTMLG. We then explored the effects of external periodic perturbations, such as circularly polarized light (CPL) and waveguide-generated linearly polarized light (WGL). CPL breaks chiral symmetry, creating a gap that distorts the Hofstadter spectrum. These distortions are more pronounced for asymmetric twist configurations. In contrast, WGL preserves chiral symmetry and has a tunable, non-monotonic effect on the bandwidth. This makes WGL a reliable tool for engineering electronic properties. These results demonstrate how (1+3+1)-CTMLG combines the effects of light-matter interactions with moiré physics. This allows accurate control of the electronic properties and fractal spectra by adjusting external fields and twist angles.

\end{abstract}
\maketitle
\section{Introduction}
Twisted bilayer graphene (TBLG) gives rise to 
a two-dimensional moiré superlattice formed by stacking two graphene layers with a small relative twist angle
\cite{cao2018unconventional, cao2018correlated}. At specific twist angles, particularly near the magic angle of approximately $\theta = 1.05^\circ$, the electronic band structure is significantly altered by the moiré pattern, resulting in the formation of essentially flat bands \cite{bistritzer2011moire}. These flat bands suppress the kinetic energy of electrons, making electron-electron interaction dominate to give rise to strongly correlated phenomena \cite{bistritzer2011moire, cao2018unconventional, cao2018correlated}. Furthermore, studies have shown that unconventional superconductivity can appear close to the magic angle, accompanied by correlated insulating states and magnetic phases \cite{bistritzer2011moire, cao2018unconventional, cao2018correlated, stepanov2020untying, cao2021nematicity, benlakhouy2022chiral}. 
Twisted trilayer graphene (TTLG) has become an important platform for studying moiré-induced electronic phenomena. When three graphene layers are stacked with a small twist, a long-wavelength moiré pattern is formed, which significantly changes the band structure and electronic interactions \cite{cao2018unconventional, liu2014evolution, cao2016superlattice, benlakhouy2023floquet}. TTLG has attracted attention for hosting highly correlated and topologically non-trivial phases \cite{li2019electronic}, having nearly flat topological bands near magic angles \cite{po2018origin, yuan2018model, hejazi2019multiple}, and exhibiting unique optical properties \cite{zuo2018scanning, mora2019flatbands, wu2018theory}. It has been widely investigated theoretically and experimentally \cite{park2021tunable, hao2021electric, assi2021floquet}.


The study of moiré physics in TBLG and TTLG has paved the way for exploring more complex multilayer graphene systems. Notably, double-twisted multilayer graphene (DTMLG) has emerged as an important extension of double-twisted trilayer graphene (DTTLG) \cite{cao2018unconventional, cao2018correlated, jiang2019charge, yankowitz2019tuning, kerelsky2019maximized, popov2023magic, guo2018pairing, wu2018theory, liu2018chiral, zhang2020correlated, chaudhary2021quantum, lu2019superconductors, khalaf2019magic, li2020experimental, stepanov2020untying, guinea2018electrostatic, ding2023mirror, liang2022moire}. 
DTMLG replaces at least one monolayer with a multilayer graphene $(n \geq 2)$ layer, which significantly alters interlayer interactions and broadens the spectrum of potential moiré band structures 
\cite{khalaf2019magic, carr2020ultraheavy, shin2021stacking, lei2021mirror, wu2021lattice, ding2023mirror, liang2022moire}.
DTMLG can be divided into two types: alternating twisted multilayer graphene (ATMLG) and chirally twisted multilayer graphene (CTMLG). {This classification is based on the relative rotation direction of the twist angles.}
In ATMLG, the twist angles alternate in sign $(\theta,-\theta, \theta,\cdots)$. In CTMLG, all twist angles are in the same direction. This key difference results in unique moiré band structures and electronic properties \cite{wang2019new, lee2019theory}. DTMLG structures are labeled as (X+Y+Z)-DTMLG, where $X$, $Y$, and $Z$ indicate the number of layers in the bottom, middle, and top vdW layers, respectively. 
Recent studies have examined various DTMLG configurations, such as (1+3+1)-DTMLG and (2+1+2)-DTMLG. These configurations show different electronic behaviors due to variations in the central vdW layer 
\cite{ding2023mirror, liang2022moire}.
In (1+3+1)-DTMLG, the middle vdW layer is Bernal-stacked trilayer graphene, while the top and bottom layers are monolayer graphene (MLG). In contrast, (2+1+2)-DTMLG has bilayer graphene (BLG) as the top and bottom vdW layers, with monolayer graphene in the middle. These stacking differences affect the interlayer coupling, moiré band structures, and correlated phases \cite{ding2023mirror, liang2022moire}. Among these systems, the (1+3+1)-CTMLG configuration stands out due to its enhanced stability and mirror symmetry decomposition. This allows it to be split into subsystems of opposite parity \cite{ding2023mirror}, producing unique moiré band structures with tunable flat bands and correlated phases. {Building on recent findings and our previous work \cite{benlakhouy2022chiral, benlakhouy2023floquet}, we investigate the electronic properties of (1+3+1)-CTMLG, in particular under the presence of external magnetic fields and periodic driving lights, to explore its potential for hosting novel quantum states.}

The electronic band structures in graphene-based systems, including (1+3+1)-CTMLG, can be significantly affected by periodic time-dependent external driving fields
\cite{benlakhouy2022chiral, seo2019ferromagnetic, oka2019floquet, oka2009photovoltaic, luo2021tuning, kibis2016magnetoelectronic}.
High-frequency electromagnetic fields near the Dirac point have been shown to modify the energy spectrum, depending on their polarization \cite{kunold2020floquet}. For example, linear polarization produces an anisotropic gapless spectrum, while circular polarization opens an isotropic energy gap \cite{suarez2013electronic}. These alterations also impact the transport properties of TBLG and multilayer graphene. They become highly dependent on the polarization of electromagnetic fields. In the case of (1+3+1)-CTMLG, the interplay between its unique chirally twisted configuration and external driving fields provides a robust system for exploring tunable transport and electrical properties \cite{wang2024correlated}. Manipulating light intensity and polarization is much easier than adjusting twist angles. 
This makes light a practical tool for tuning the properties of advanced moiré systems.
The integration of Floquet theory into the study of electromagnetic interactions with materials has gained considerable attention, providing a robust framework for studying time-periodic systems. To simplify the analysis, various methods have been developed to convert these time-dependent problems into equivalent time-independent forms \cite{benlakhouy2022chiral, rodriguez2021low, abanin2017effective, itin2015effective, mikami2016brillouin}. Recently, the combination of Floquet theory with moiré physics has opened up new possibilities for characterizing and understanding topological phases in moiré materials \cite{li2010observation, assi2021floquet, vogl2020floquet, rodriguez2020floquet, lu2021valley}.

Recent studies have investigated the effects of two different types of light-circularly polarized light (CPL) and transverse magnetic (TM) waveguide light (WGL)-on TBLG, revealing important changes in its electronic energy spectrum
\cite{vogl2019analog, vogl2020floquet}. 
In the case of CPL, a rotating frame Hamiltonian has been developed that applies to both weak and strong fields in the high-intermediate frequency regimes \cite{vogl2019analog}. This Hamiltonian, designed for scenarios where the conventional Van Vleck (vV) approximation breaks down, provides a more accurate representation of quasi-energies in such systems \cite{rodriguez2021low}. Using this framework, we extend these findings to (1+3+1)-CTMLG to investigate how its chirally twisted configuration and multilayer structure respond to external periodic driving fields.
This could open new ways to study Floquet-engineered phenomena and the resulting electronic properties.

More specifically, we intend to extend the effective Hamiltonian approach \cite{vogl2020effective, vogl2020floquet}, previously applied to the study of TBLG \cite{bistritzer2011moire}, to the CTMLG system \cite{ding2023mirror, liang2022moire}. Building on our earlier investigations of TBLG under various external driving conditions \cite{benlakhouy2022chiral}, we explore how the unique chirally twisted configuration and multilayer structure of CTMLG influence its electronic and transport properties. This work focuses on the Hofstadter butterfly phenomenon \cite{hofstadter1976energy}, where electrons exposed to a magnetic field exhibit a fractal energy spectrum. This remarkable pattern has been observed in various systems. Examples include monolayer graphene (MLG) \cite{hunt2013massive, ponomarenko2013cloning}, AB-stacked bilayer graphene (AB-BLG) on hexagonal boron nitride (hBN) \cite{dean2013hofstadter}, as well as in square, honeycomb, triangular \cite{oh2000energy, oh2006energy}, and kagome lattices \cite{du2018floquet}. It has also been studied in TBLG \cite{bistritzer2011moire, benlakhouy2022chiral, chou2020hofstadter, saito2021hofstadter, kim2023replica} and more recently in TTLG configurations \cite{benlakhouy2023floquet, imran2023hofstadter}, highlighting the interplay between moiré superlattices and external fields. Previous studies have examined the influence of periodic driving (light) on the Hofstadter butterfly in systems such as MLG under a magnetic field combined with laser. The study of the effect of light on the Hofstadter butterfly in multilayer systems such as TBLG and TTLG remains relatively limited \cite{wang2015atomic, wang2009quantum, lawton2009spectral, wang2013kicked, lababidi2014counter, zhou2014floquet, kooi2018genesis, wackerl2019driven, ding2018quantum}. In particular, these studies have often emphasized the fractal properties of Landau levels in the presence of light. Here, we aim to go beyond this and investigate how light interacts with chirally twisted structures such as (1+3+1)-CTMLG, focusing on the intricate relationship between light, chiral symmetries, and the resulting fractal properties of the Hofstadter spectrum.

We recall that the periodicity of the moiré pattern in twisted multilayer graphene (TMLG) is closely related to the twist angles between adjacent layers. In general, the moiré pattern is periodic only for a discrete set of twist angles, such that the resulting moiré lattices from each twist are commensurate with each other. Usually, the twist angles are very small, much less than three degrees as in the present work, hence the generated moiré superlattices are very large, resulting in approximate periodicity over large areas, as can be easily seen visually. Needless to say, even the famous 1.1 degree magic angle that produced superconductivity in twisted bilayer graphene did not produce a truly periodic moiré pattern. However, the large moiré lattice that was produced was approximately periodic.

In our work, to analyze the electronic properties of  of  (1+3+1) chirally twisted graphene with different twist angles $\theta_1 \neq \theta_2$, we will use  a hybrid modeling approach based on the continuum Bistritzer-MacDonald (BM) model originally developed for twisted bilayer graphene (TBLG) \cite{bistritzer2011moire}. Due to the incommensurate nature of the double twist configuration, a global Bloch band structure cannot be defined. To overcome this, we treat each twisted interface (top-middle and bottom-middle bilayers) independently, using separate moiré Brillouin zones, and compute spectral features by truncating the momentum space couplings to a finite shell, consistent with previous studies on incommensurate multilayer systems \cite{Koshino2015, Carr2019}. For the Hofstadter butterfly calculations, which require a well-defined unit cell to apply a uniform magnetic field, we approximate the incommensurate twist angles with close rational values, thereby constructing a large periodic supercell \cite{Moon2012}. The magnetic field is incorporated via Peierls substitution, allowing flux quantization and the observation of field-dependent miniband structures. This approach captures the essential physics of quasi-periodic moiré systems under magnetic fields, as demonstrated in other studies of multilayer graphene and moiré superlattices \cite{Wu2019}.

The present work is organized as follows. In Sec. \ref{MODEL AND METHODS}, we present the theoretical framework for the (1+3+1)-CTMLG system and discuss key equilibrium properties. We focus on how symmetric and asymmetric twist angles affect the electronic band structure. In Sec. \ref{Equilibrium properties: interlayer hoppings and the Hofstadter butterfly}, we investigate the role of hopping processes in shaping the Hofstadter butterfly spectrum and show how they determine the fractal energy structure and symmetry properties for $E=0$. In Sec. \ref{Circulary polarized light}, we analyze the effects of circularly polarized light and show that increasing the driving power leads to significant enlargement and deformation of the Hofstadter butterfly, with asymmetry amplifying these effects. In Sec. \ref{Waveguide light}, we consider the influence of longitudinal waveguide light and observe the non-monotonic dependence of the bandwidth on the driving power and observe the preserved reflection symmetry around $E=0$. Finally, in Sec. \ref{SUMMARY}, we summarize our results, highlighting how light-induced effects, combined with the unique chirality and twist angle asymmetry affect the (1+3+1)-CTMLG system.


\section{theoretical background}
\label{MODEL AND METHODS}
\begin{figure}[h!]
	\vspace{0.cm}
	\centering
\subfloat[]{\includegraphics[width=0.45\linewidth]{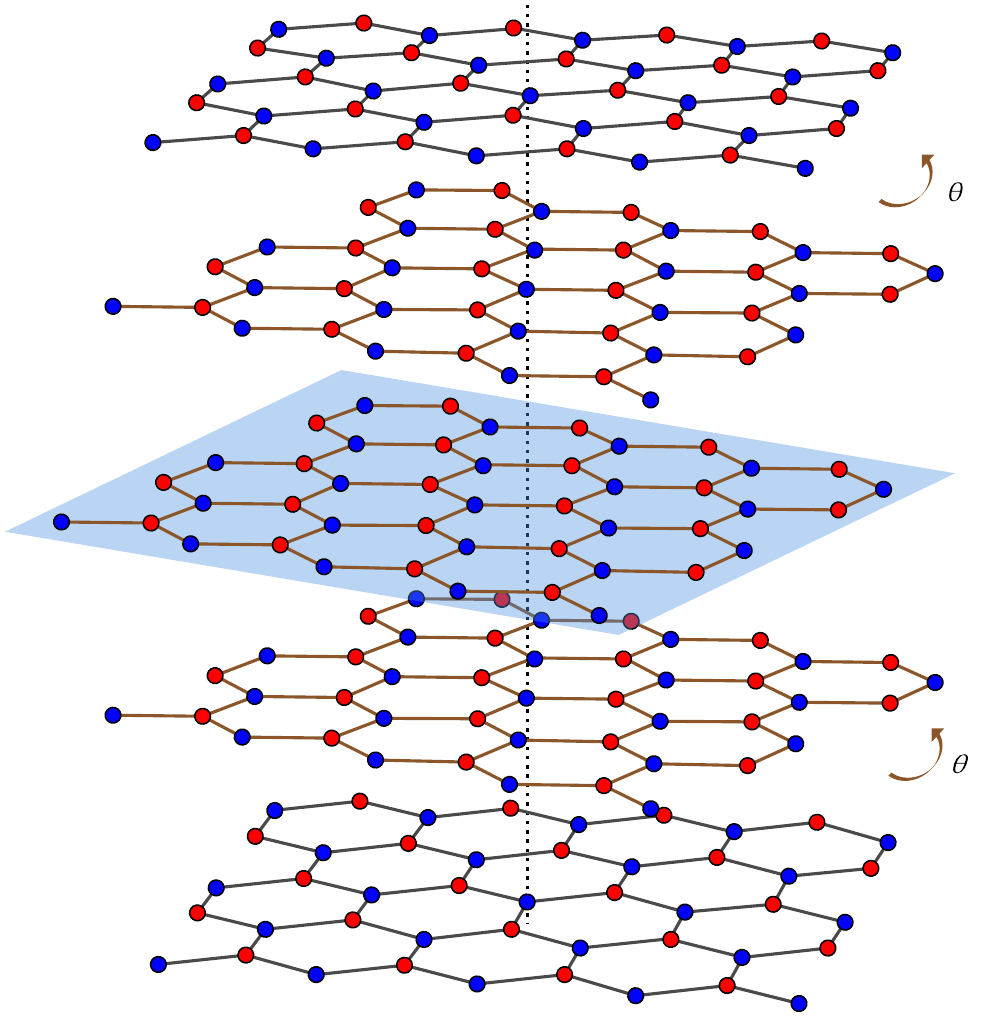}}
\subfloat[]{\includegraphics[width=0.45\linewidth]{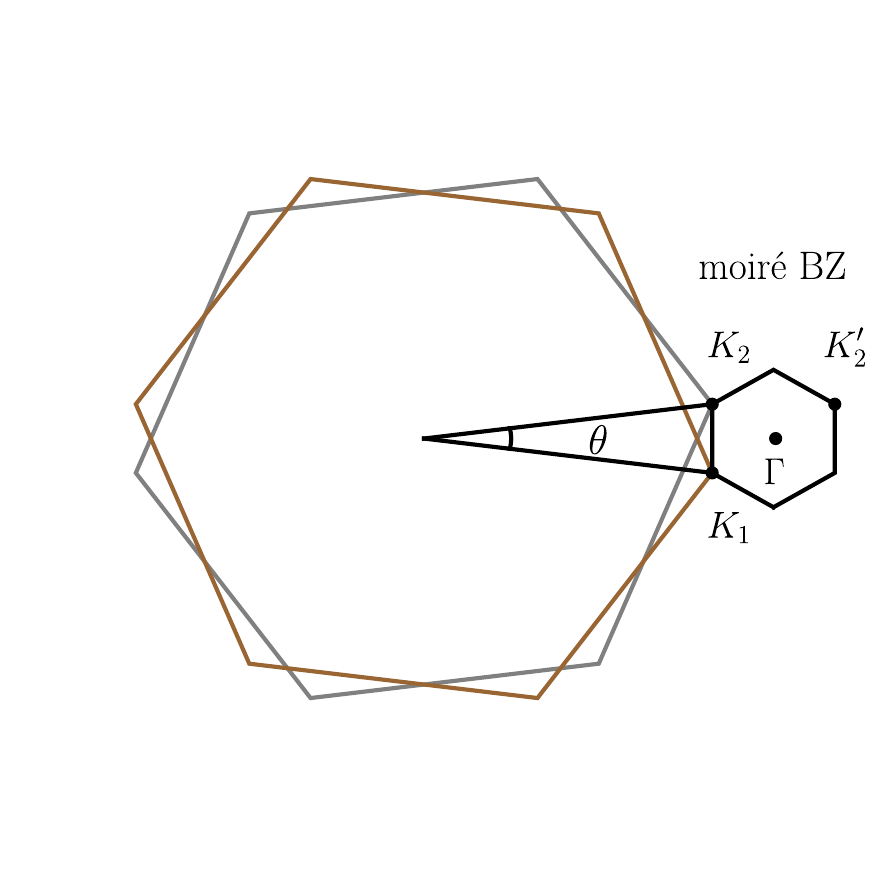}}	
	\caption{(a) The (1+3+1) CTMLG in the case where  the twist angles $\theta_1$ and $\theta_2$ have the same rotation direction ($\theta_1=\theta_2=\theta$). (b) The moiré Brillouin zone for each rotation.}\label{Schematicsystem}
\end{figure}

We begin our study by defining double twisted multilayer graphene (DTMLG) in the (1+3+1) configuration with twist angles $\theta_1$ and $\theta_2$, as illustrated in Fig. \ref{Schematicsystem}. The Hamiltonian for the DTMLG system, based on a specific basis representation, is given by \cite{bistritzer2011moire, moon2013optical, ding2023mirror}
\begin{equation}
H=\begin{pmatrix}
		{ H_1\left(\frac{\theta_1}{2}\right)} & T_{12}^{\dagger} & 0 & 0 & 0 \\
		T_{12} & {H_2\left(-\frac{\theta_1}{2}\right)} & T_{23}^{\dagger} & 0 & 0 \\
		0 & T_{23} & H_3 & T_{34}^{\dagger} & 0 \\
		0 & 0 & T_{34} & { H_4\left(\frac{\theta_2}{2}\right)} & T_{45}^{\dagger} \\
		0 & 0 & 0 & T_{45} & {H_5\left(-\frac{\theta_2}{2}\right)}
\end{pmatrix},\label{Hamiltonian131}
\end{equation}
and each  graphene layer $L$ is governed by the intralayer Hamiltonian $H_{L} $
(in unit $\hbar=1$)
\begin{equation}
	H_{L}(\theta_{i})=-iv_{\text{F}}(\boldsymbol{\sigma}_{\theta_i}\cdot\boldsymbol{\nabla}),\quad
  H_{3}=-iv_{\text{F}}(\boldsymbol{\sigma}\cdot\boldsymbol{\nabla}),  
\end{equation}
where the Fermi velocity is $v_{\text{F}} = 10^{6}$ \ \text{m/s}, and $\boldsymbol{\sigma}_{\theta_i}$ denotes the Pauli matrices acting on the sublattice space and is defined as
\begin{equation}
	\boldsymbol{\sigma}_{\theta_i}=\left(\sigma_x\cos\theta_i-\sigma_y\sin\theta_i,
	\sigma_x\cos\theta_i+\sigma_y\sin\theta_i\right).
\end{equation}
In Eq. (\ref{Hamiltonian131}), the off-diagonal blocks correspond to the interlayer hopping terms. Indeed, we have
\begin{align}
	T_{12}(\boldsymbol{r})=\sum_{j=1}^{3} e^{-i \boldsymbol{q}_{1j} \cdot \boldsymbol{r}} T_{j},\quad T_{45}(\boldsymbol{r})=\sum_{j=1}^{3} e^{-i \boldsymbol{q}_{2j} \cdot \boldsymbol{r}} T_{j},
\end{align}
where $\boldsymbol{q}_{i1}=k_{\theta_i}(0, -1)$ and $\boldsymbol{q}_{i2,3}=k_{\theta_i}(\pm\sqrt{3},1)$ are the nearest neighbor vectors of the moiré Brillouin zone (MBZ).
{These vectors are scaled by $k_{\theta_i}=2k_D\sin\frac{\theta_i}{2}$, where $k_D=4\pi/3a_0$ denotes the Dirac momentum and $a_0=2.46 \textup{~\AA}$ is the graphene lattice constant.}
The $T_j$ matrices describe the tunneling between sublattices, which depends on the stacking type. 
They are \cite{li2019electronic}
\begin{align}
	&	T_{j}^{\mathrm{AB}}=[T_{j}^{\mathrm{BA}}{ }]^{\dagger}=\begin{pmatrix}
			\omega_0	e^{i \frac{2 j \pi}{3}} &\omega_1 \\
			\omega_1	e^{-i \frac{2 j \pi}{3}} & \omega_0e^{i \frac{2 j \pi}{3}}
		\end{pmatrix},
    \label{Eq: interlayer-hoppinf-matrices}
\\
	&	T_{j}^{\mathrm{AA}}=\begin{pmatrix}
			\omega_0 & \omega_1e^{-i \frac{2 j \pi}{3}} \\
			\omega_1e^{i \frac{2 j \pi}{3}} & \omega_0
		\end{pmatrix},
	\end{align}
and $\omega_i$ represent the relaxation due to AB and BA stacking, which are more energetically favorable than AA-stacked regions \cite{vogl2020floquet, li2020floquet, katz2020optically}. In addition, $\text{AA}$ and $\text{AB}$ regions provide different interlayer lattice constants \cite{guinea2019continuum}. In our case, we account for these variables by setting $(\omega_0,\omega_1)=(0.9~\omega_1,110~\text{meV})$ as the $(\text{AB/BA}, \text{AA})$ type hopping amplitude. They are similar to those in TBLG, where distortions can be seen in the relaxed lattice when next-neighbor-layer interactions are excluded \cite{assi2021floquet, vogl2020floquet, katz2020optically}. Finally, the interlayer hopping between the 1st and 2nd graphene layers (no twist) is
\begin{equation}
T_{2,3}=\left(\begin{array}{cc}
	0 & t_{\perp} \\
	0 & 0
\end{array}\right),
\end{equation}
with  $T_{3,4}=T_{2,3}^{\dagger}$. These mathematical tools provide a framework for achieving our goal, which is to study the effects of two external factors: a perpendicular magnetic field and two different forms of light.
Specifically, we will analyze how DTMLG responds to these influences by examining its energy levels, considering both identical and different twist angles. This study aims to provide deeper insights into the behavior of twisted systems and their unique properties.

\section{Equilibrium analysis}
\label{Equilibrium properties: interlayer hoppings and the Hofstadter butterfly}


In Appendix \ref{appendix A}, we show how a magnetic field can be included in the Hamiltonian Eq. \eqref{Hamiltonian131} describing (1+3+1)-CTMLG in the equilibrium case.
We then present numerical results that reveal the main features of interlayer hopping and the Hofstadter butterfly spectrum. These results provide important insights into the underlying physics of interlayer hopping and the Hofstadter butterfly spectrum. For this, we analyze both symmetric and asymmetric twist angle configurations and demonstrate their different effects on the electronic structure.
Our study shows how the magnetic field and interlayer interactions work together to shape the behavior of the system. This approach not only deepens our understanding of twisted graphene systems but also highlights the fascinating quantum effects that occur in these materials.
Here, we would like to stress that the periodicity of the Moiré pattern in twisted multilayer graphene (TMLG) is tightly linked to the twist angles between adjacent layers. In general, the Moiré pattern is periodic only for a discrete set of twist angles so that the resulting moiré lattices from each twist are commensurate with each other. Usually twist angles are very small, much less than three degrees as it is the case in the present work, hence the generated Moiré superlattices are very large resulting in an approximate periodicity over large areas as it is visually recognized. Needless to mention that even the famous 1.1-degree magic angle that produced superconductivity in twisted bilayer graphene did not produce a truly periodic Moiré pattern. However, the large Moiré lattice that was generated was approximately periodic.

\subsection{Symmetric case ($\theta_{1} = \theta_{2}$)}
\begin{figure}[htb]
	\vspace{0.cm}
	\centering
\includegraphics[width=0.5\linewidth]{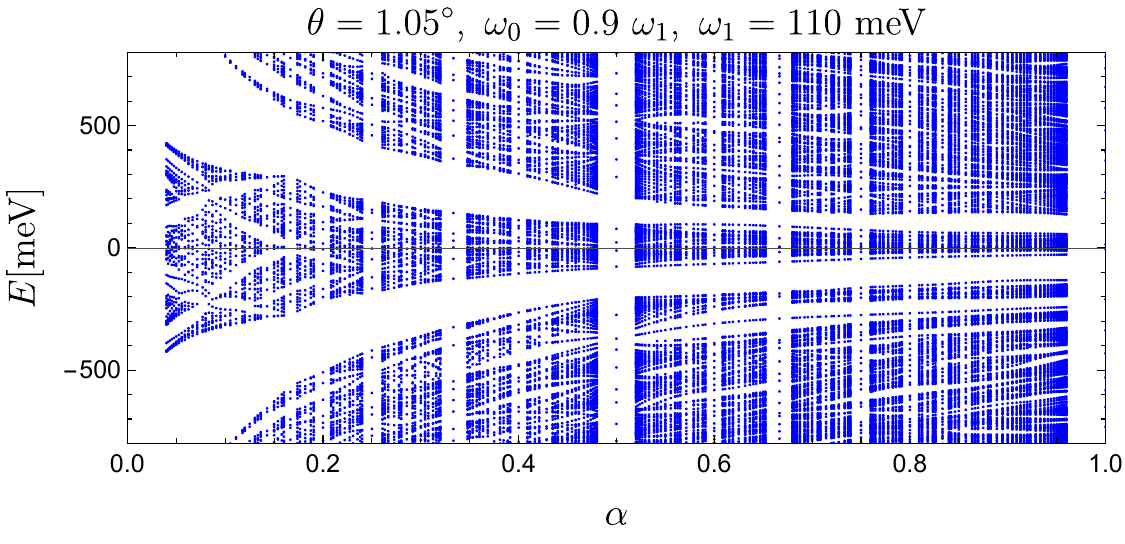}\includegraphics[width=0.5\linewidth]{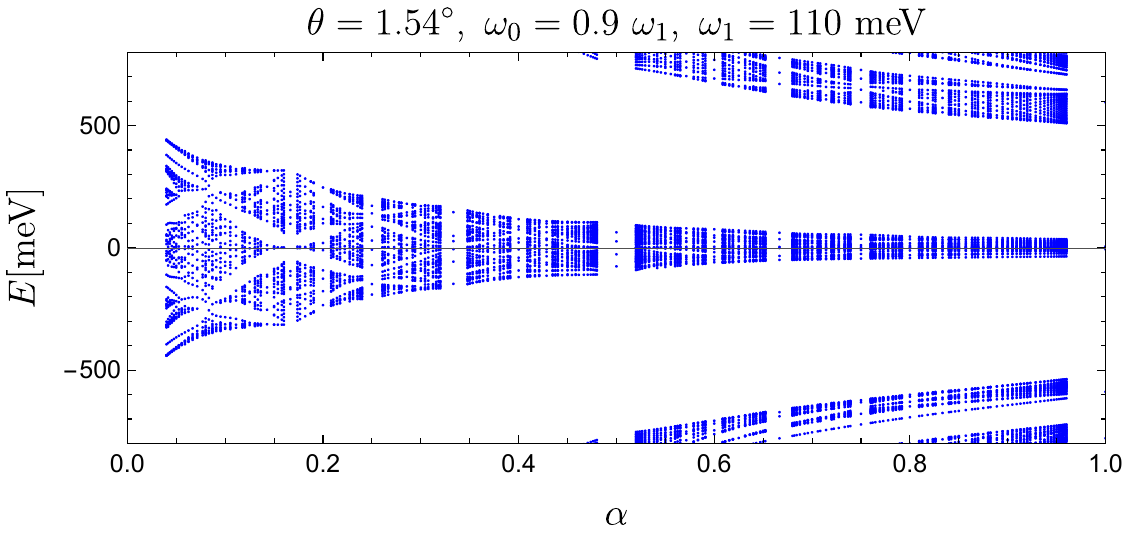}\\
	\includegraphics[width=0.5\linewidth]{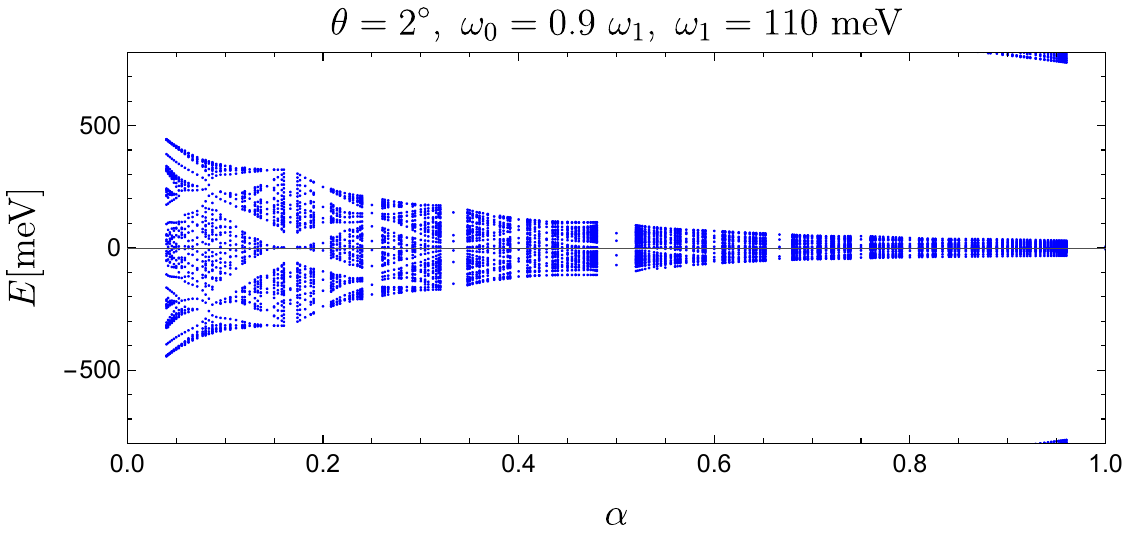}\includegraphics[width=0.5\linewidth]{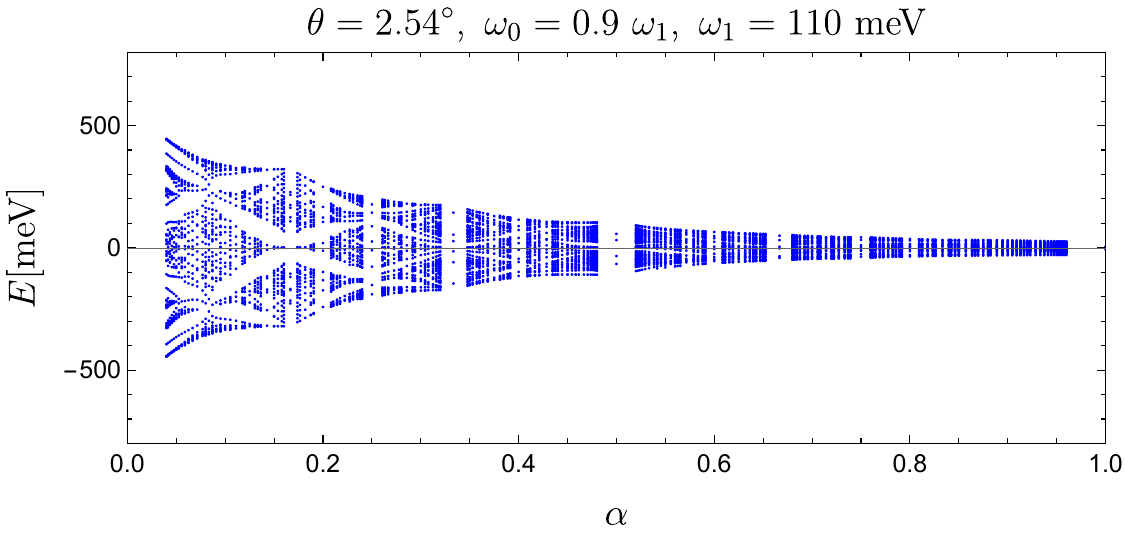}	
	\caption{The Hofstadter butterfly in $(1+3+1)$-CTMLG submitted to $B$  as a function of the magnetic flux ratio $\alpha=\frac{p}{q}$ for the symmetric case.
    }\label{equilibrium-case1}
\end{figure}

To show how the Hofstadter butterfly spectrum changes with twist angles ($\theta_{1}, \theta_{2}$) and magnetic field ($B$), we plot the symmetric case $\theta_{1} = \theta_{2}=\theta $ in Fig. \ref{equilibrium-case1}, with $\theta = 1.05^\circ$, $1.54^\circ$, $2^\circ$, and $2.54^\circ$.
For the tunneling parameters and $(\omega_0, \omega_1)=(0.9~\omega_1,110~\text{meV})$, the results show distinct transitions in the spectrum as $\theta$ varies. For $\theta = 1.05^\circ$, the spectrum exhibits a well-defined fractal structure with clear sub-band gaps and also prominent Landau levels over the full range of flux ratios $\alpha$. 
This shows the strong influence of the moiré superlattice potential on the system at small twist angles.
The interaction between interlayer coupling and magnetic effects is particularly significant. These results are consistent with the results in \cite{shi2021exotic} and highlight how these interactions control the behavior of the system. As the twist angle increases to reach the magic angle $\theta = 1.54^\circ$, there are noticeable changes in the butterfly spectrum. Higher energy Landau levels shift upward, making the fractal structure less noticeable, while the central $n=0$ Landau level remains dominant. This is consistent with the expected behavior associated with this magic angle. Here, the interlayer tunneling induces flat bands and leads to enhanced electronic correlations and reduced fractal features at higher energies \cite{cao2018correlated}.
At twist angles $2^\circ$ and $2.54^\circ$, the inter-layer interaction between graphene layers increases, leading to a mixing of the Landau levels and a blurring of the higher energy bands.
This increased interaction reduces the distinct features of these bands, however, the central region, associated with the $n=0$ Landau level, retains small energy gaps.
This behavior indicates a transition from strongly correlated electronic states at small twist angles to more dispersive band structures at larger angles; this observation is consistent with the findings in the study Landau levels in twisted bilayer graphene and semiclassical orbits, which discusses how different twist angles affect the mixing of Landau levels and the resulting electronic spectrum in TBLG \cite{hejazi2019landau}.

\subsection{Asymmetric case ($\theta_{1}\neq \theta_{2}$)}

\begin{figure}[htb]
	\vspace{0.cm}
	\centering
	\includegraphics[width=0.5\linewidth]{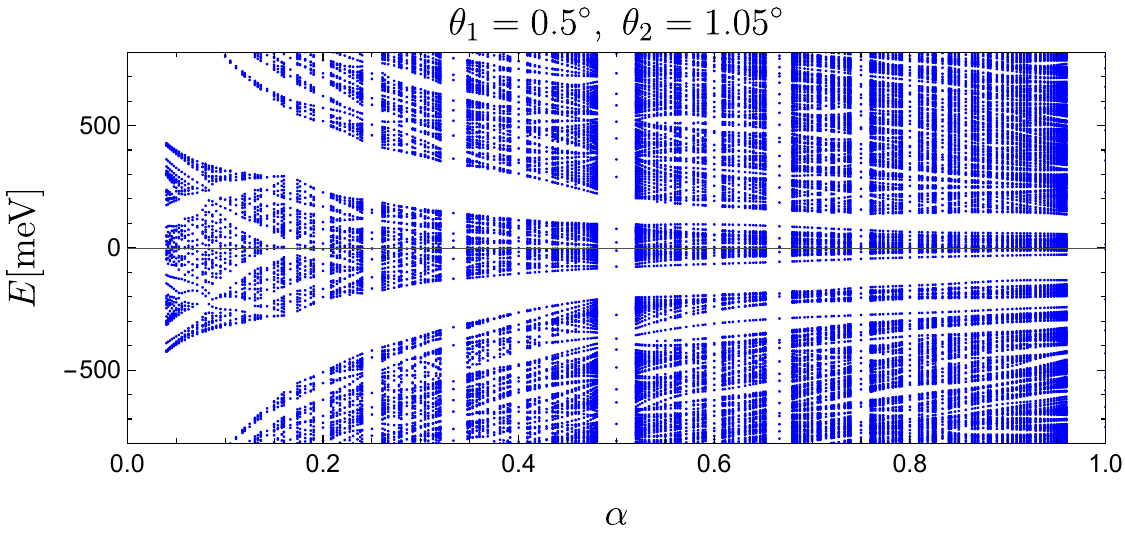}\includegraphics[width=0.5\linewidth]{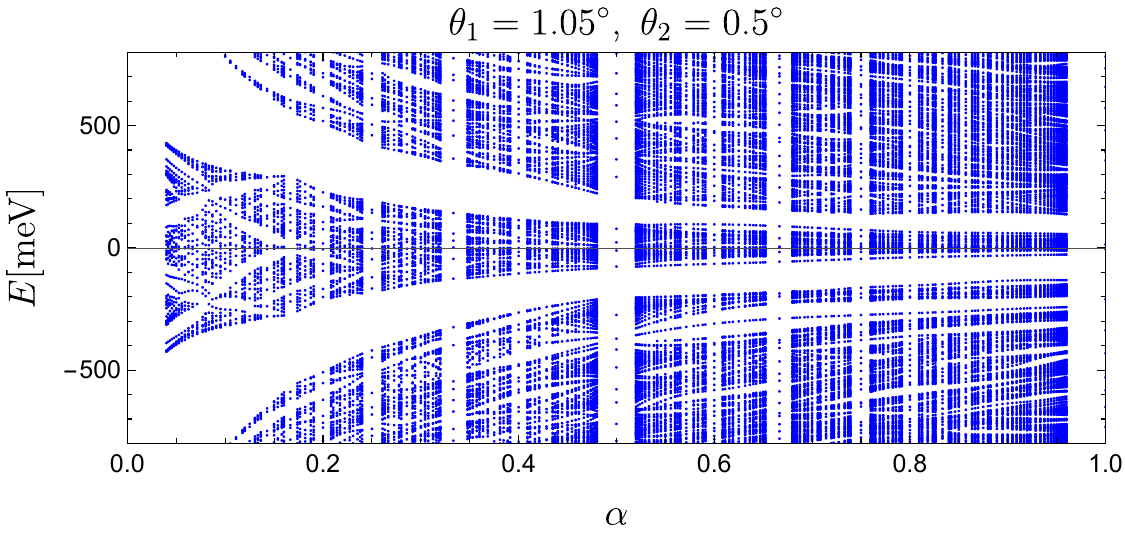}
	\caption{The Hofstadter butterfly in
		$(1+3+1)$-CTMLG subject to $B$ as a function of the magnetic flux ratio $\alpha=\frac{p}{q}$ for the asymmetric case.
	}\label{equilibrium-case2}
\end{figure}

In Fig. \ref{equilibrium-case2}, we present the asymmetric case and investigate how two different twist angles influence the Hofstadter butterfly spectrum. Considering the configurations $\theta_{1} = 0.5^\circ, \theta_{2} = 1.05^\circ $ and the swapped one, $\theta_{1} = 1.05^\circ, \theta_{2} = 0.5^\circ $, the spectra exhibit highly fractal features with sharp Landau levels and well-defined sub-band gaps, driven by the dominant moiré effects of the smaller angles. surprisingly, the results show that the Hofstadter butterfly spectra are identical for both configurations. This equivalence can be attributed to the mirror symmetry of the (1+3+1) DTMLG system along the $z$ axis, where swapping $\theta_{1}$ and $\theta_{2}$ mirrors the structure without changing its electronic properties \cite{suarez2013electronic}. In (1+3+1)-CTMLG, the relative magnitudes of the twist angles $\theta_1$ and $\theta_2$ are the only parameters that control the interlayer coupling strengths, rather than their order \cite{zhu2020twisted}. 
It is worth mentioning that exchanging the twist angles $\theta_1 \leftrightarrow \theta_2$ does not affect the effective moiré potentials or the superlattice patterns.
As a result, the Hamiltonian of the system remains the same under this exchange, leading to identical electronic spectra for both configurations. This symmetry highlights a fundamental property of the (1+3+1) structure: while the twist angle asymmetry $\theta_1 \neq \theta_2$ affects the spectrum, the sequence of twist angles does not. This invariance simplifies the theoretical analysis and enables a unified description of the electronic properties. It remains unaffected by the twist sequence \cite{bistritzer2011moire}.

\subsection{Interlayer hopping}

\begin{figure}[htb]
	\vspace{0.cm}
	\centering
	\includegraphics[width=0.5\linewidth]{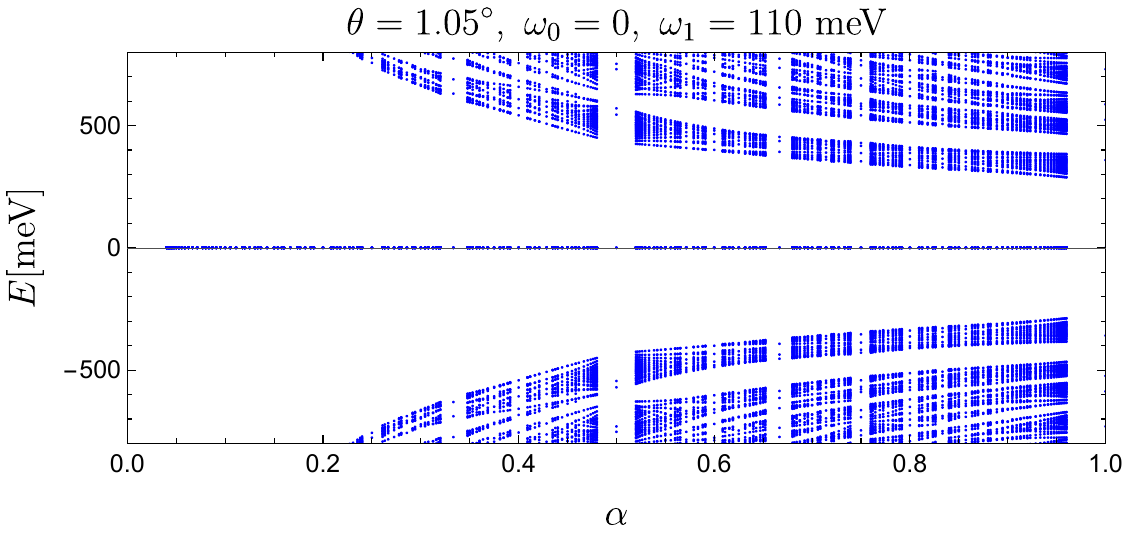}\includegraphics[width=0.5\linewidth]{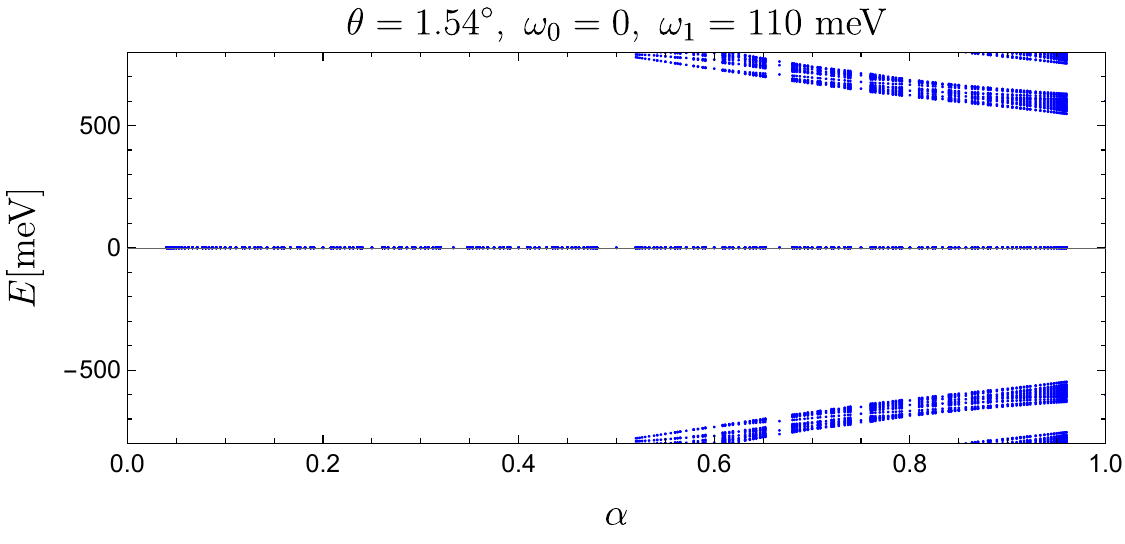}
	\includegraphics[width=0.5\linewidth]{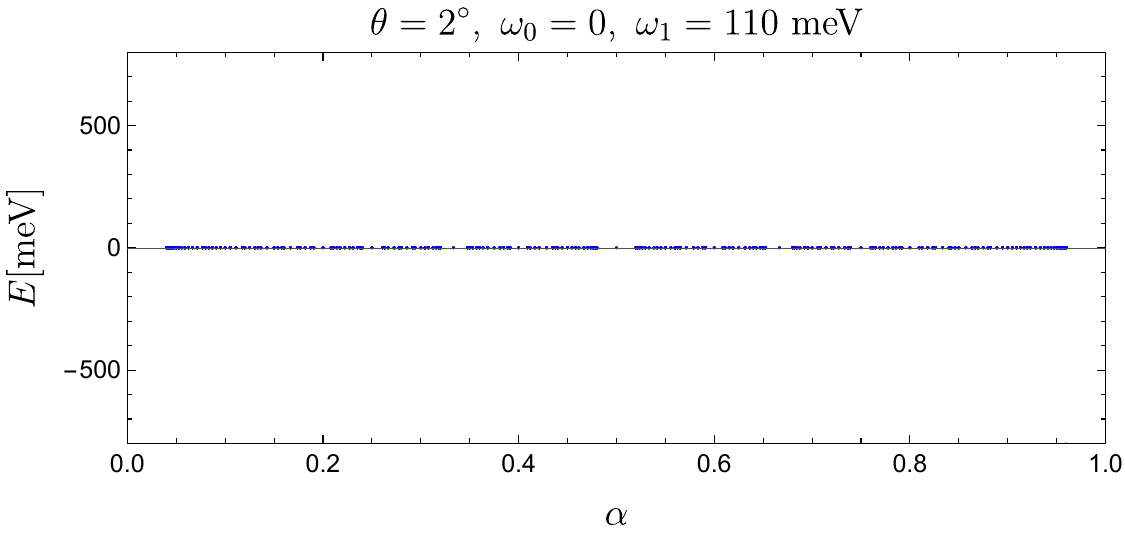}\includegraphics[width=0.5\linewidth]{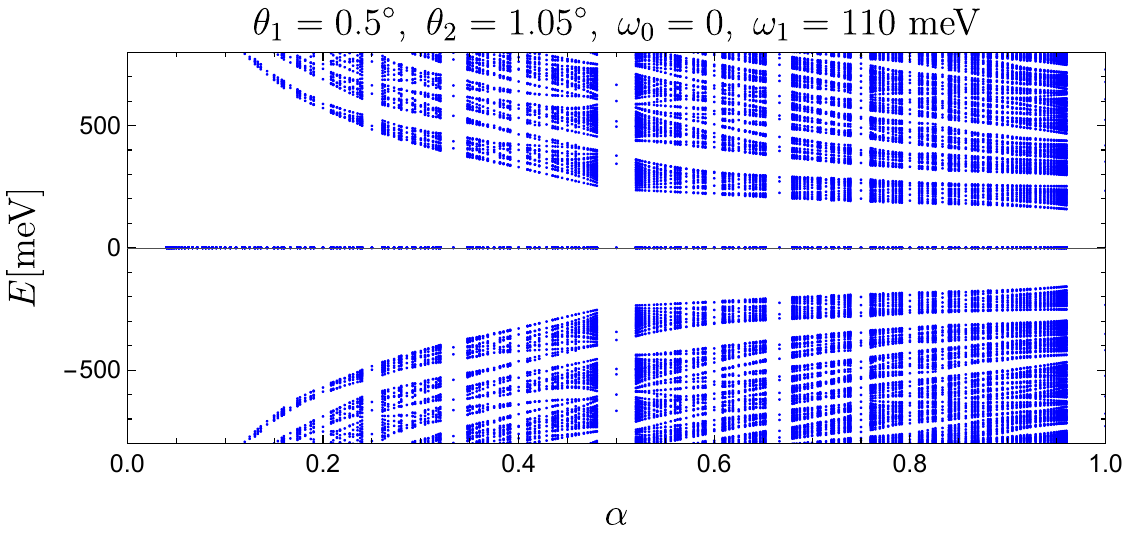}
	\caption{The Hofstadter butterfly in $(1+3+1)$-CTMLG a function of the magnetic flux ratio $\alpha=\frac{p}{q}$ for the interlayer hoppings $\omega_0=0$, and~$\omega_1=110$ meV.}\label{equilibrium-case3}
\end{figure}

For our (1+3+1) DTMLG structure, the next step is to investigate how the interlayer hopping amplitudes, specifically the AA-type hopping $ \omega_0$ and AB/BA-type hopping $\omega_1 $, influence the Hofstadter butterfly spectrum \cite{nam2017lattice, li2020floquet}. We aim to determine which type of hopping is most critical to the spectrum's structure. In particular, we explore the case where $\omega_0 = 0 $ and $ \omega_1 = 110 ~\text{meV}$, corresponding to the chiral model. In this configuration, the central part of the Hofstadter butterfly spectrum is expected to collapse into a zero-energy line \cite{benlakhouy2022chiral, benlakhouy2023floquet}, as the lowest Landau level of graphene resides on a single sublattice. This distinct characteristic is a result of $\omega_1$ couples sublattices, leaving the lowest Landau level unaffected while significantly impacting higher-energy levels that involve both sublattices.
In Fig. \ref{equilibrium-case3}, the Hofstadter butterfly spectrum in the chiral model shows a strong influence of interlayer hopping and twist angles on the structure of the spectrum. For $\theta= (1.05^\circ, 1.54^\circ, 2^\circ)$, the spectrum collapses into a zero-energy line, as the lowest Landau level resides on a single sublattice while higher-energy states are suppressed due to the absence of AA-type hopping. We also consider the asymmetric case, where $\theta_1 = 0.5^\circ, \theta_2 = 1.05^\circ$ (see Fig. \ref{equilibrium-case3}), small sub-band features appear at higher energies, but the zero-energy line is still dominant. For smaller twist angles, the moiré potential is enhanced, leading to well-defined fractal features across the spectrum. These results demonstrate that small twist angles enhance the fractal-like structure, while larger angles $\theta > 2^\circ$ lead to more dispersive bands. In addition, while asymmetry in the twist angles $\theta_1 \neq \theta_2$ introduces minor variations in the sub-band distribution, it does not significantly alter the spectrum under the chiral model.

\section{Circularly polarized light}
\label{Circulary polarized light}

We study the combined effect of circularly polarized light (CPL) and magnetic field on the Hofstadter butterfly spectrum in $(1+3+1)$-CTMLG. We begin by constructing the corresponding Hamiltonian and then proceed to numerical analysis.

\subsection{Effective Hamiltonian}
We assume a CPL is transmitted perpendicular to the DTMLG at frequency $\Omega$ and driving strength $ A $. 
Eventually, the effect of light enters the Hamiltonian through minimal substitution \cite{vogl2020effective, assi2021floquet, dehghani2015out}
\begin{align}
	&k_{x} \rightarrow k_{x}(t)=k_{x}-A \cos (\Omega t),\\
	& k_{y} \rightarrow {k}_{y}(t)= k_{y}-A \sin (\Omega t).
\end{align}
As a result, we obtain a time-periodic Hamiltonian, $H(\mathbf{x}, \mathbf{k},t) = H(\mathbf{x}, \mathbf{k},t+2\pi/\Omega)$. It is well known that such a Hamiltonian can be accurately treated using an effective time-independent Floquet Hamiltonian \cite{vogl2020effective}. In the present case, a numerically advantageous parameterization will be beneficial. Therefore, let us look at how to obtain an effective time-dependent description and what additional physical parameters are introduced.
Transforming a periodically driven Hamiltonian into a rotating frame (RF) is a non-perturbative method. This allows to obtain an effective time-independent Hamiltonian via a unitary transformation 
$U(t)$ \cite{vogl2020effective}
\begin{equation}
	H_R =U^{\dagger}(t)\left(H-i \partial_{t}\right) U(t).
\end{equation}
A Hamiltonian is generated by subsequent time averaging if a suitable frame is chosen. This is a more reliable description compared to the van Vleck or Floquet-Magnus approximations, which are common high-frequency estimators \cite{rodriguez2021low}. It has been shown that using a well-chosen unitary transformation \cite{vogl2020effective, assi2021floquet}, for TTLG subjected to CPL, a highly accurate effective Hamiltonian can be derived as
\begin{widetext}
	\begin{equation}
H=\begin{pmatrix}
			H_1 & \tilde{T}_{12}^{\dagger} & 0 & 0 & 0 \\
			\tilde{T}_{12} & H_2 & \tilde{T}_{23}^{\dagger} & 0 & 0 \\
			0 & \tilde{T}_{23} & H_3 & \tilde{T}_{34}^{\dagger} & 0 \\
			0 & 0 & \tilde{T}_{34} & H_4 & \tilde{T}_{45}^{\dagger} \\
			0 & 0 & 0 & \tilde{T}_{45} & H_5
		\end{pmatrix}\label{Hamiltonian of 131}
	\end{equation}
\end{widetext}
where each $H_L$ is given by
\begin{equation}
H_L=	v_{\mathrm{RF}} R(\theta_{L}) \mathbf{k}\cdot \boldsymbol{\sigma}-\Delta_{\mathrm{RF}} \sigma_{3},
\end{equation}
and $R(\theta_L)$ denotes the rotation matrix. The Fermi velocity has been changed and is now equal to
\begin{equation}
	v_{\mathrm{RF}}=v_{\mathrm{F}} J_{0}\left(-\frac{6 \gamma}{\Omega} J_{1}\left(\frac{2 A a_{0}}{3}\right)\right) J_{0}\left(\frac{2 A a_{0}}{3}\right),
\end{equation}
with $J_0$ is the first kind of zeroth Bessel function. Light also causes the creation of a band gap, which is expressed as
\begin{equation}
	\Delta_{\mathrm{RF}}=-\frac{3 \gamma}{\sqrt{2}} J_{1}\left(\frac{2 A a_{0}}{3}\right) J_{1}\left(-\frac{6 \sqrt{2} \gamma}{\Omega} J_{1}\left(\frac{2 A a_{0}}{3}\right)\right).
\end{equation}
The interlayer tunneling matrices Eq. \ref{Eq: interlayer-hoppinf-matrices} are also modified. If we express  $T_{j}$ as $T_{j}=\sum_i T_{j,i}\sigma_i$, where $T_{j,n}$ are expansion coefficients, 
we get new hopping matrices
$\tilde{T}_{j}$ as
\begin{equation}
	\tilde{T}_{j}^{\mathrm{\text{AB}}}=[\tilde{T}_{j}^{\mathrm{BA}}{ }]^{\dagger}=\sum_i T_{j,i}\tilde \sigma_i ,
\end{equation}
with 
\begin{align}
	&\tilde{\sigma}_{1,2}=J_{0}(\nu) \sigma_{1,2},\\
	&\tilde{\sigma}_{0,3}=\sigma_{0,3}+\left(J_{0}(\sqrt{2} \nu)-1\right)\left[\sigma_{0,3} \sin ^{2}\frac{\theta}{2}\pm\frac{i}{2} \sigma_{3} \sin \theta\right],
\end{align}
where $ \nu=(-6\gamma /\Omega) J_{1}\left(2 A a_{0} / 3\right)$, and $\sigma_{1,2,3}$ are the Pauli matrices, and $\sigma_0$ is the  identity matrix. 

\subsection{Numerical analysis}
\begin{figure}[htb]
	\centering
	\includegraphics[width=0.5\linewidth]{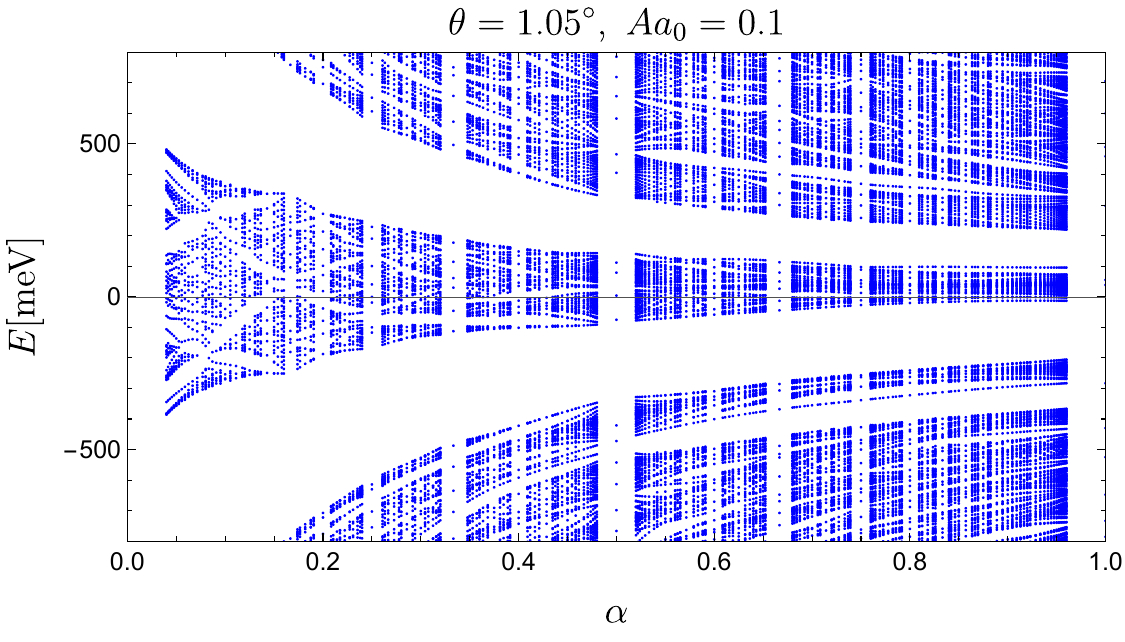}\includegraphics[width=0.5\linewidth]{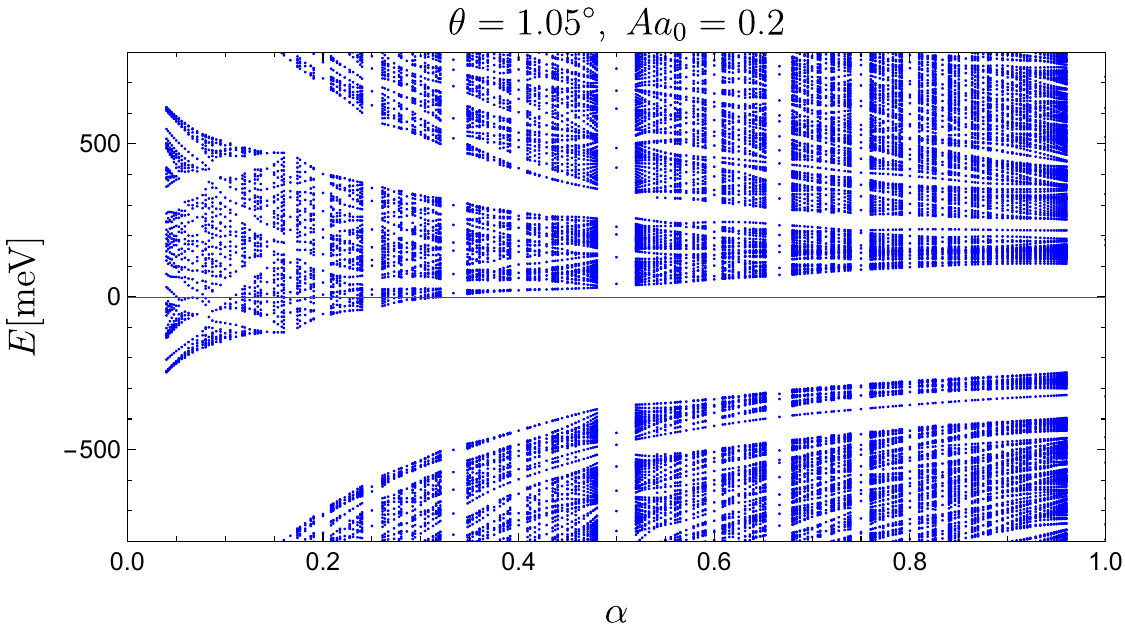}\\
	\includegraphics[width=0.5\linewidth]{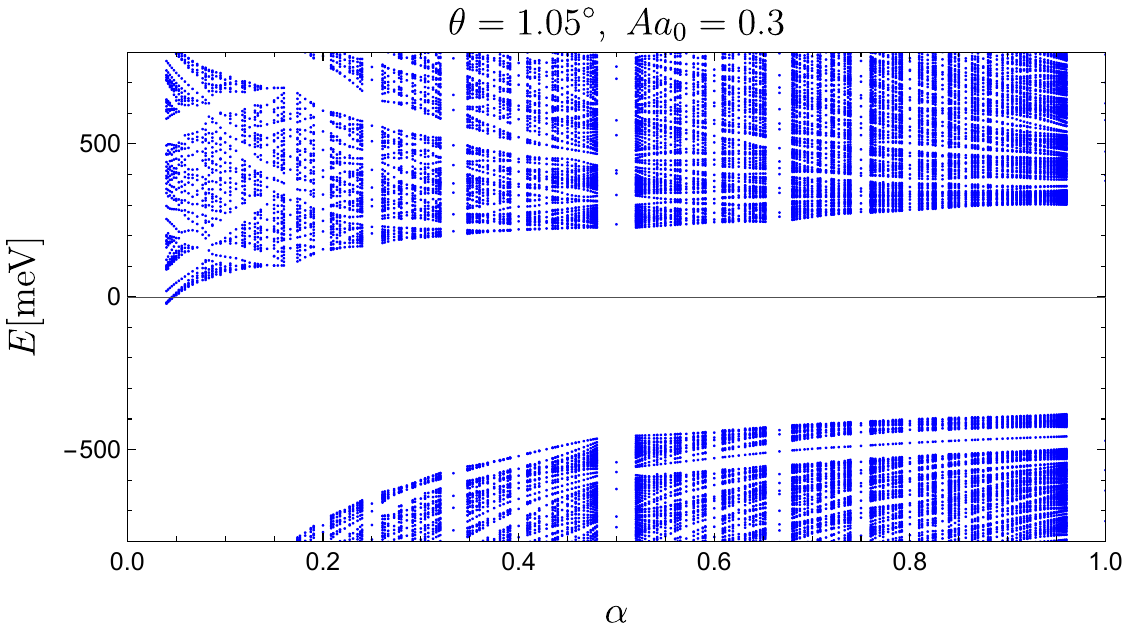}\includegraphics[width=0.5\linewidth]{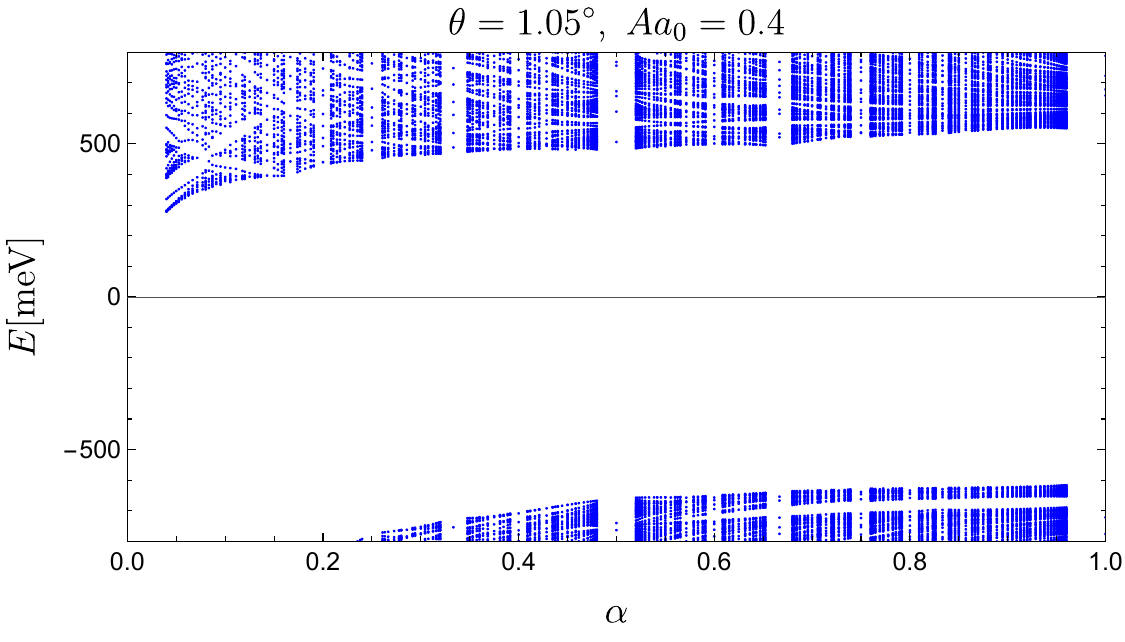}
	\caption{The Floquet Hofstadter butterfly spectrum in (1+3-1)-CTMLG subject to $B$ and CPL as a function of the magnetic flux ratio $\alpha= \frac{p}{q}$ for the symmetric case, with
      $\gamma=2364$ meV and $\theta=1.05^{\circ}$.}\label{CPL1}
\end{figure}

\begin{figure}[htb]
	\centering
	\includegraphics[width=0.5\linewidth]{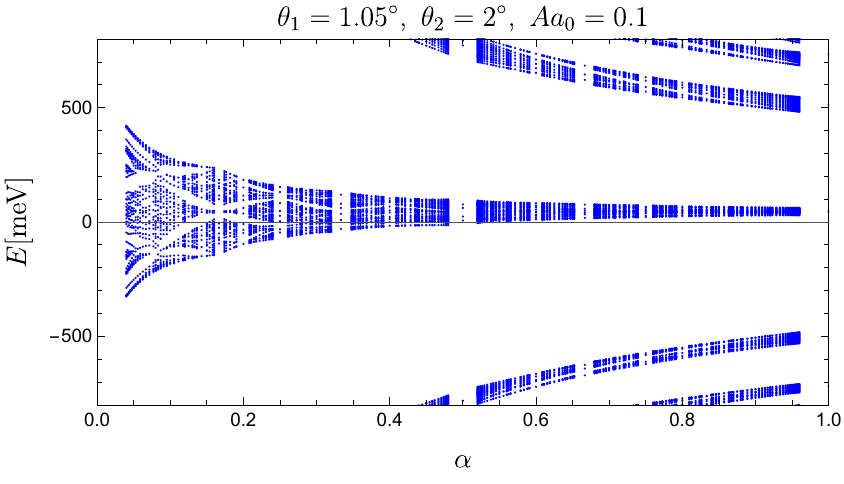}\includegraphics[width=0.5\linewidth]{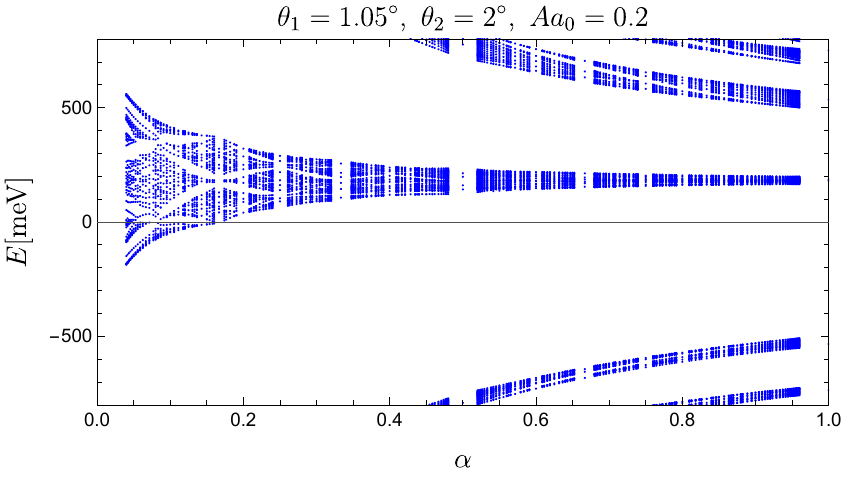}\\
		\includegraphics[width=0.5\linewidth]{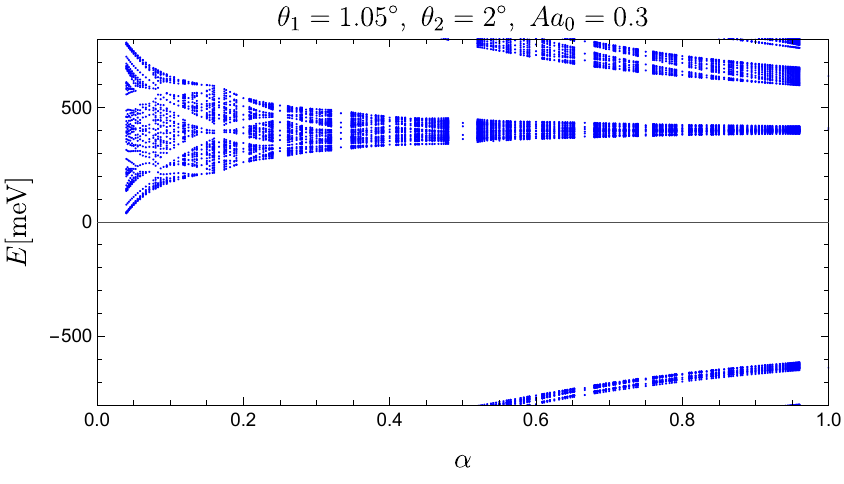}\includegraphics[width=0.5\linewidth]{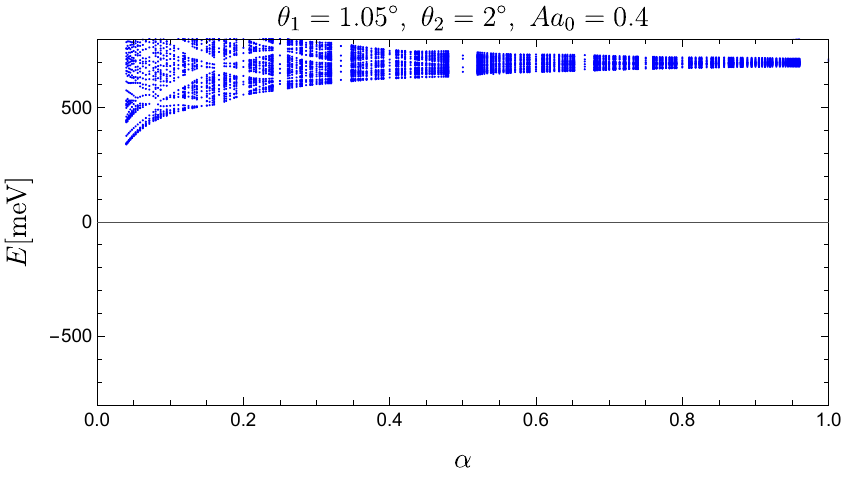}
	\caption{The Floquet Hofstadter butterfly spectrum in (1+3-1)-CTMLG as a function of the magnetic flux ratio $\alpha= \frac{p}{q}$ for the asymmetric case, with $\gamma=2364$ meV and $\theta_1=1.05^{\circ}, \theta_2=2^{\circ}$.}\label{CPL2}
\end{figure}

Under circularly polarized light (CPL), the Hofstadter butterfly spectrum in TBLG exhibits increasing asymmetry at $E=0$ as the driving strength increases, due to the breaking of chiral symmetry by the $\Delta_{\text{RF}}$ term.
This asymmetry mainly affects the central branch, corresponding to the $n=0$ Landau level, which shifts upward to higher energies. In contrast, higher-energy levels retain some symmetry under certain conditions, as illustrated in Figs. \ref{CPL1} and \ref{CPL2}. In our (1+3+1) DTMLG results, we observe similar Floquet-induced asymmetry and distortions, especially at higher driving strengths. For the symmetric case, $\theta = 1.05^\circ$ in Fig. \ref{CPL1}, the spectrum maintains some uniformity at lower driving strengths but becomes increasingly distorted at higher driving strengths. For the asymmetric case $\theta_1=1.05^\circ, \theta_2=2^\circ$ in  Fig. \ref{CPL2}, asymmetry amplifies these distortions, especially in the central region, as the interaction between the two twisted interfaces introduces additional effects on the Floquet dynamics.
Furthermore, the central band, which is typically shifted upward under right-handed circularly polarized light, can be shifted downward using left-handed circularly polarized light. This occurs because, to leading order in Floquet theory, $\Delta_{\text{RF}}$ reverses sign $\Delta_{\text{RF}} \rightarrow -\Delta_{\text{RF}}$, changing the direction of the shift. This demonstrates the tunability of the Hofstadter butterfly spectrum through the handedness of circularly polarized light.

\section{Linearly Polarized light}

In this section, we consider another form of light in addition to the magnetic field and analyze its impact on the energy levels. More precisely, we consider a waveguide linearly polarized light and identify the parameters that affect the physics of the present system.

\label{Waveguide light}
\begin{figure}[htb]
	\centering
	\includegraphics[width=0.5\linewidth]{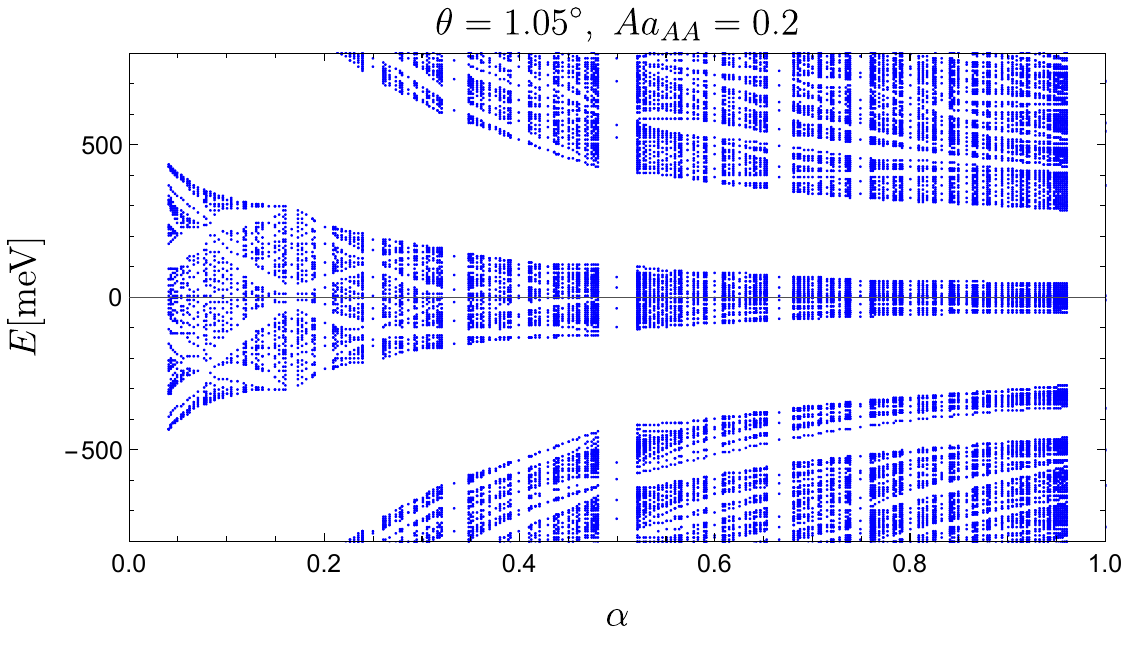}\includegraphics[width=0.5\linewidth]{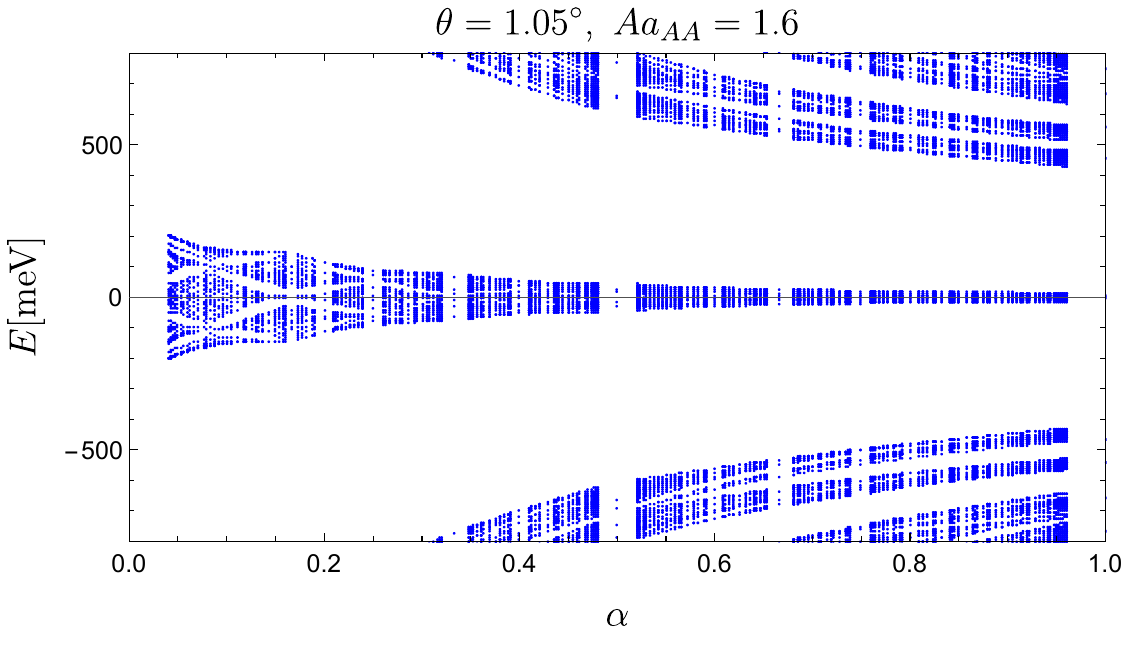}\\
	\includegraphics[width=0.5\linewidth]{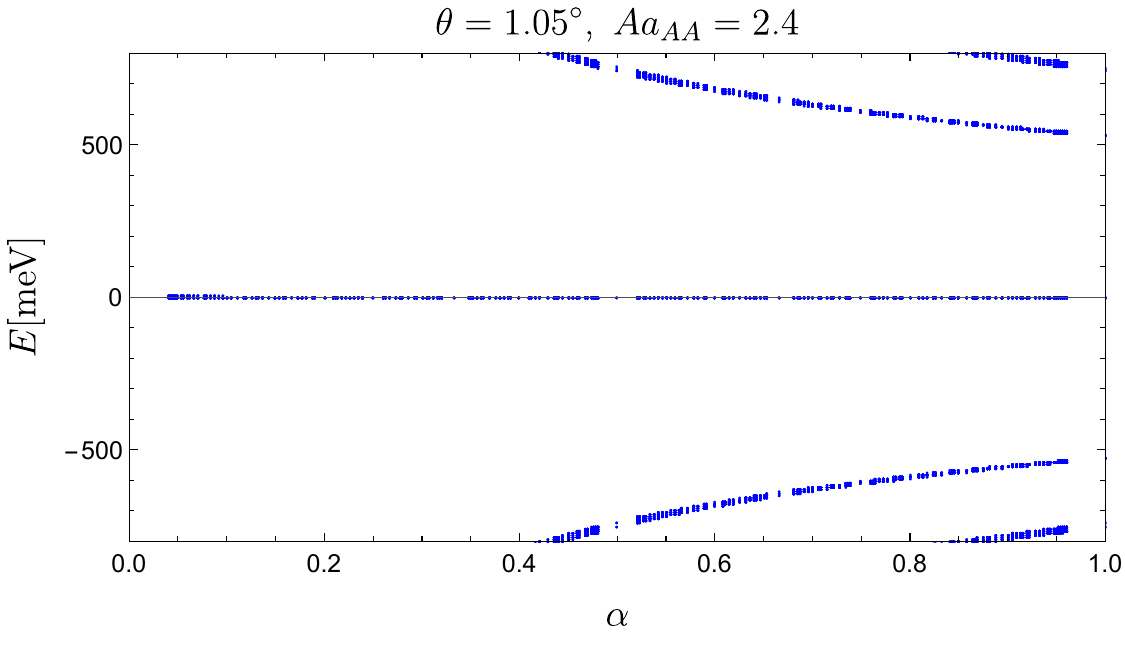}\includegraphics[width=0.5\linewidth]{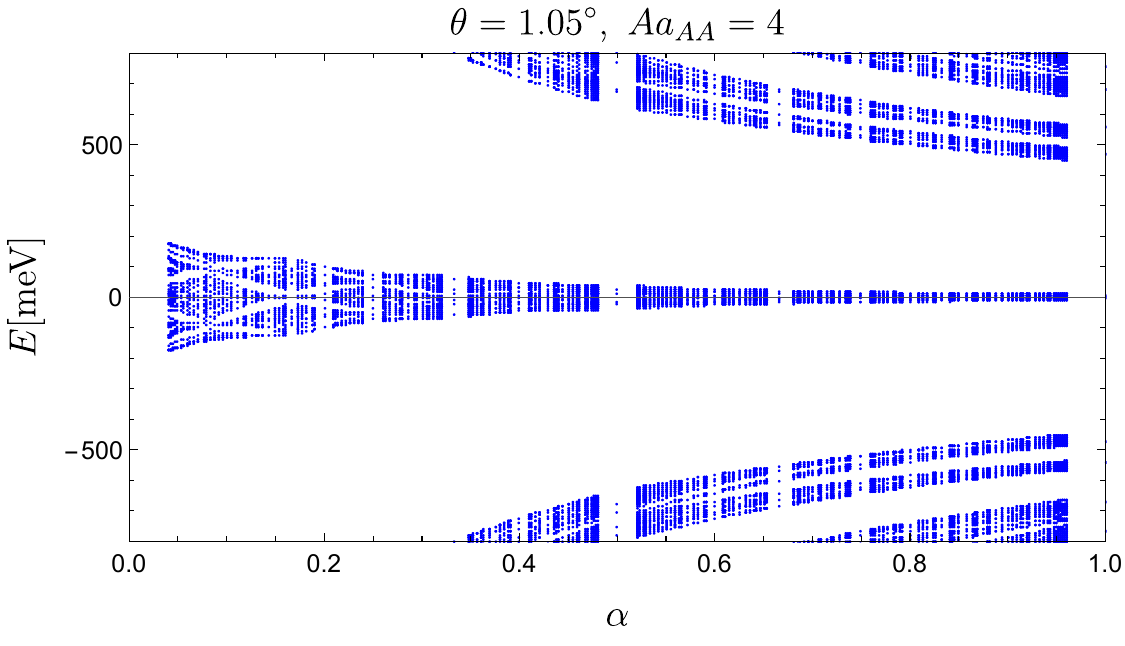}
	\caption{The Floquet Hofstadter butterfly spectrum as a function of $\alpha=\frac{p}{q}$. The representative driving strengths are chosen from $Aa_{AA}=0.2$ to $Aa_{AA}=4$. The parameters used are $\gamma=2364$ meV and $\theta=1.05^{\circ}$.}\label{WGL1}
\end{figure}
\begin{figure}[htb]
	\centering
	\includegraphics[width=0.5\linewidth]{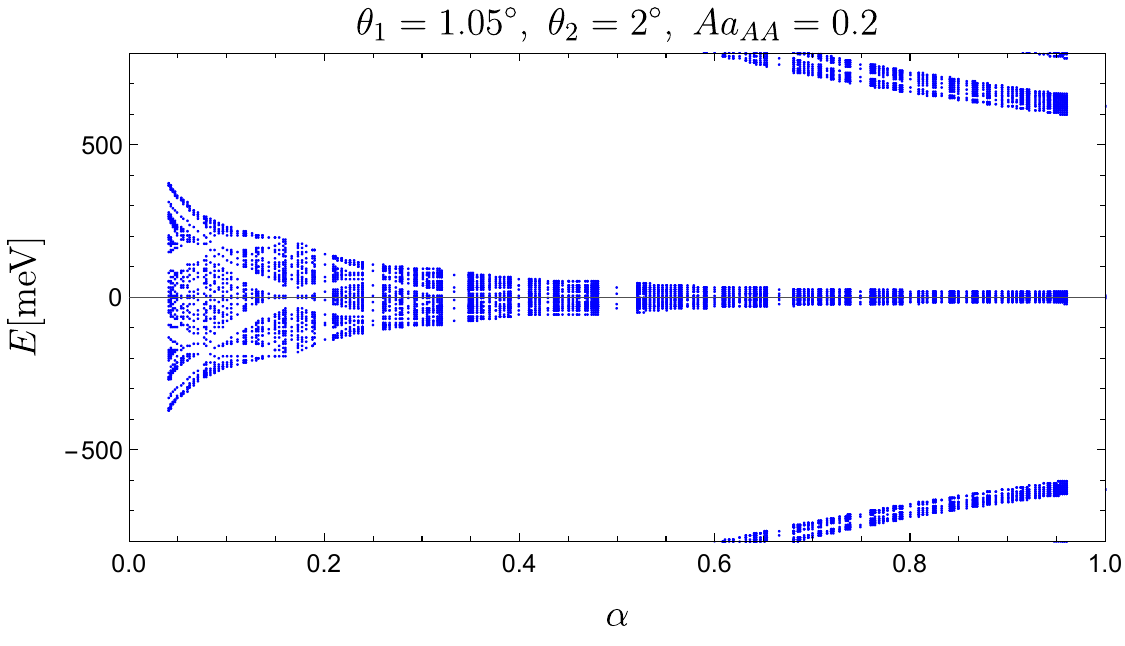}\includegraphics[width=0.5\linewidth]{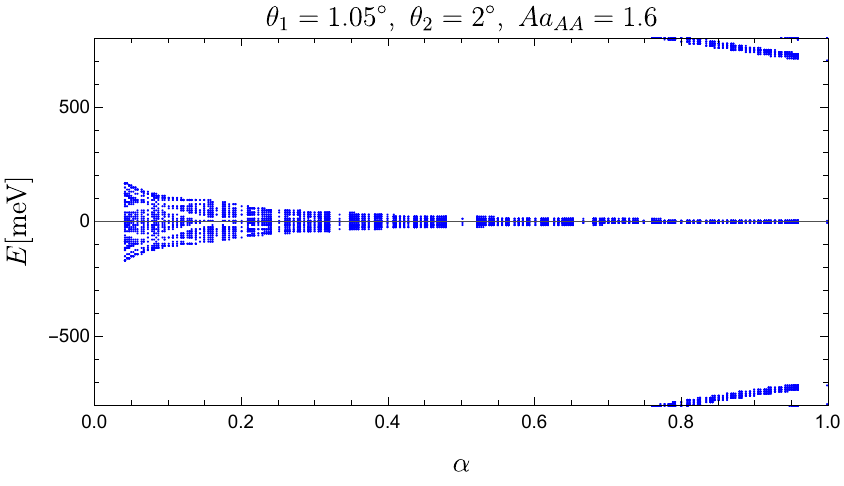}\\
	\includegraphics[width=0.5\linewidth]{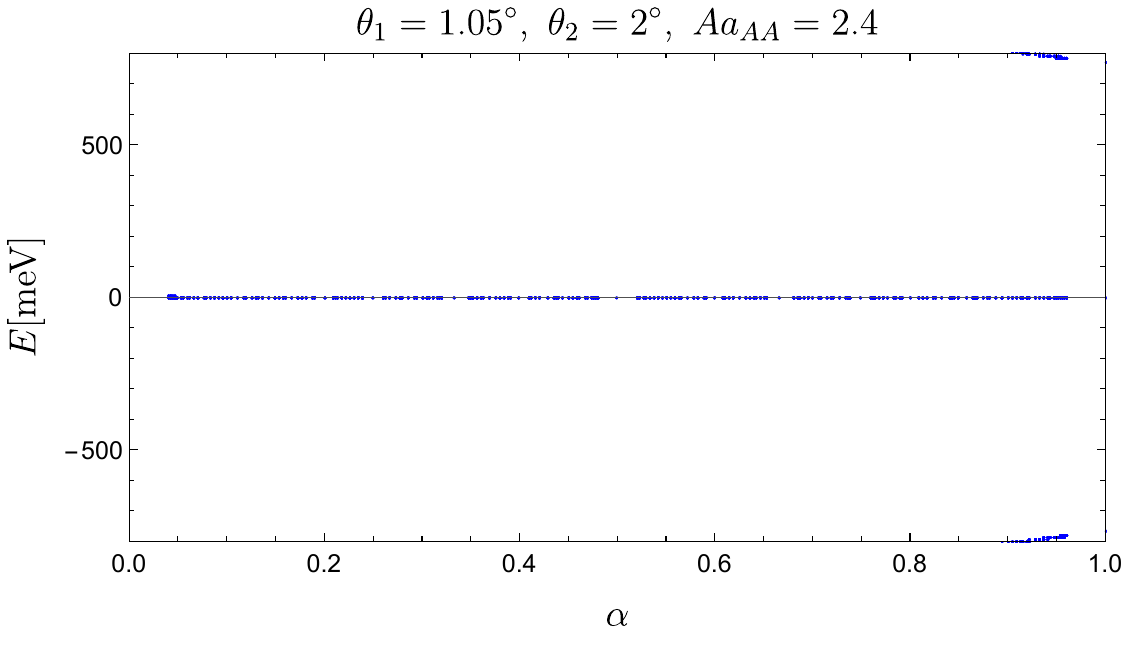}\includegraphics[width=0.5\linewidth]{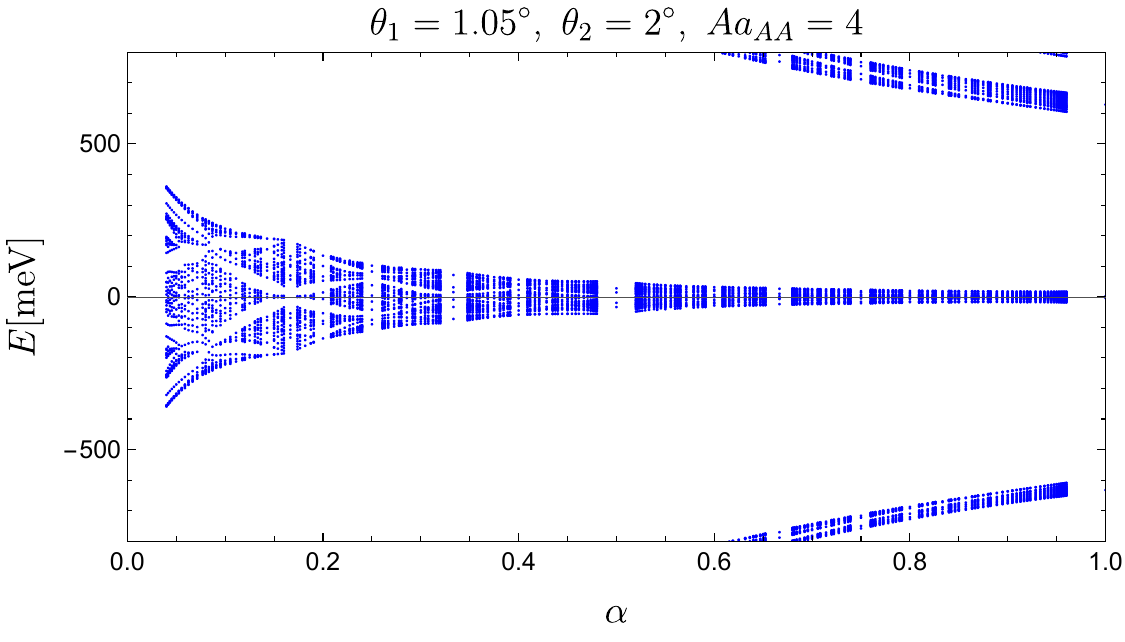}
	\caption{The Floquet Hofstadter butterfly spectrum as a function of $\alpha=\frac{p}{q}$. The representative driving strengths are chosen from $Aa_{AA}=0.2$ to $Aa_{AA}=4$. The parameters used are $\gamma=2364$ meV and $\theta_1=1.05^{\circ}$, $\theta_2=2^{\circ}$.}\label{WGL2}
\end{figure}

\subsection{Theoretical model}
We will look now at a second type of light: longitudinal light coming from a waveguide. In this case, the boundary conditions of the waveguide allow light to have longitudinal components, described by $\mathbf{A}=\operatorname{Re}\left(e^{i k_{z} z-i \Omega t}\right) \mathbf{e}_{z}$, even under vacuum. For a deeper understanding of how this is derived, see \cite{vogl2020floquet} or standard electromagnetism texts such as \cite{jackson1999classical}. It is important to note that this effect can only be replicated at the tight-binding level using Peierls substitution. That is 
\begin{equation}
t_{ij} \rightarrow t_{ij} e^{ \left(-\int_{{r}_{i}}^{r_{j}} \mathbf{A}\cdot d\mathbf{l}\right)}.	
\end{equation}
Here \(\mathbf{A}\) represents the vector potential. In the continuum model, the Hamiltonian hopping terms are associated with \(\omega_i\), which is transformed as follows
\begin{equation}
w_{i} \rightarrow w_{i} e^{ \left(-\int_{r_{i}}^{r_{j}} \mathbf{A}\cdot d\mathbf{l}\right)}.
\end{equation}	
This effect can be captured in the high-frequency regime of the continuum model by adjusting the interlayer couplings. Specifically, the couplings \(w_1\) and \(w_0\) are modified as follows
\begin{equation}
	\begin{aligned}
		&w_{1} \rightarrow \tilde{w}_{1}=J_{0}\left(\left|Aa_{A B} \right|\right) w_{1}, \\
		&w_{0} \rightarrow \tilde{w}_{0}=J_{0}\left(\left|Aa_{A A} \right|\right) w_{0},
	\end{aligned}
	\label{eq:wireplacements}
\end{equation}
\noindent where $a_{\text{AA}} = 0.36$ nm and $a_{\text{AB}} = 0.34$ nm are the interlayer distances in the $\text{AA}$ and $\text{AB}$ regions, respectively. The function $J_0$ denotes the zero-order Bessel function resulting from the high-frequency approximation.


\subsection{Results and discussion}

We numerically investigate the effect of waveguide light on the Hofstadter butterfly in (1+3+1) DTMLG. Following an approach similar to TBLG \cite{benlakhouy2022chiral,benlakhouy2023floquet}, we select a set of dimensionless driving strengths, \(Aa_{\text{AA}}\). It is important to note that
\begin{equation}
\frac{Aa_{\text{AB}}}{Aa_{\text{AA}}}=\frac{a_{\text{AB}}}{a_{\text{AA}}}.
\end{equation}
The Hofstadter butterfly spectrum for (1+3+1)-CTMLG under waveguide light is analyzed for symmetric twist angles $\theta = 1.05^\circ$ in Fig. \ref{WGL1}. As the driving strength $Aa_{\text{AA}}$ increases, the bandwidth exhibits non-monotonic behavior: it decreases in the range from $Aa_{\text{AA}} = 0.2$ to $2.4$, reflecting suppression of effective hopping, and then increases for $Aa_{\text{AA}} > 2.4$, indicating enhanced effective hopping. This trend is similar to that seen in TBLG. It shows that the combination of twist angles and waveguide light actively changes the electronic structure.
It is important to note that the reflection symmetry around the energy axis $E= 0$ is preserved throughout the spectrum. This symmetry preservation, absent for circularly polarized light, highlights the unique effects of waveguide light. Furthermore, no inverted bands are observed; this is consistent with the absence of symmetry-breaking terms in the high-frequency expansion for this twisted angle configuration. For the asymmetric case $\theta_1=1.05^\circ$ and $\theta_2=2^\circ$ in Fig. \ref{WGL2}, we see that the Hofstadter butterfly spectrum shows a distinct response to waveguide light compared to the symmetric case. As shown in Fig. \ref{WGL1}, the behavior remains non-monotonic. However, in Fig.~\ref{WGL2}, {the effects of waveguide light are more pronounced due to the asymmetry between the two twisted interfaces.}
Similarly to the symmetric case, the bandwidth initially decreases for driving strengths from $A a_{\text{AA}} = 0.2$ to $2.4$, indicating suppressed interlayer hopping. In the asymmetric case, however, this suppression is stronger, leading to narrower bands. Beyond $A a_{\text{AA}} = 2.4$, the bandwidth increases again, signaling enhanced hopping at higher driving strengths.
This behavior occurs in both symmetric and asymmetric cases. However, in the asymmetric case (\(\theta_1 = 1.05^\circ, \theta_2 = 2^\circ\)), the suppression is stronger and the recovery is less smooth,
{due to the imbalance between the two twisted interfaces.}

\subsection{Comparison with CPL}


The effects of circularly polarized light (CPL) and waveguide light (WGL) on the Hofstadter butterfly spectrum in (1+3+1)-CTMLG exhibits different behaviors. These differences arise from the type of light interaction along with the symmetry and strength of the interlayer coupling.
In both symmetric (\(\theta = 1.05^\circ\)) and asymmetric (\(\theta_1 = 1.05^\circ, \theta_2 = 2^\circ\)) configurations, CPL strongly perturbs the chiral symmetry. This leads to non-uniform distortions in the spectrum, especially for the central $n=0$ Landau level. As the driving power increases, this asymmetry becomes more pronounced, leading to significant changes in the mini-band structure. On the other hand, WGL maintains reflection symmetry around $E=0$, ensuring that the spectrum remains balanced between positive and negative energy states.

Another important difference is in how the bandwidth evolves under each type of light. In CPL, the spectrum is progressively deformed, and the minibands shift unevenly due to the formation of a gap caused by the breaking of chiral symmetry. In contrast, for WGL, the bandwidth changes in a non-monotonic manner. Initially, for driving strengths between $A a_{\text{AA}}$ between $ = 0.2$ and $2.4$, the bandwidth decreases, indicating that interlayer hopping is suppressed. However, for $A a_{\text{AA}} > 2.4$, the bandwidth begins to increase again, indicating that interlayer coupling becomes stronger.

These variations highlight the different effects of circularly polarized light (CPL) and waveguide light (WGL) on the system. CPL causes uneven changes in the electronic structure. This leads to irregular shifts in the energy bands and disrupts the overall symmetry of the system. In contrast, WGL preserves the inherent symmetry of the structure while inducing non-monotonic changes in the bandwidth. In other words, WGL keeps the system in balance but causes fluctuations in the band structure. These fluctuations are non-linear and lead to more complex changes in electronic properties. 
{These differences highlight the different ways in which each form of light affects the behavior of the system and provide potential avenues for tailored control of electronic properties.}




\section{SUMMARY AND CONCLUSION}
\label{SUMMARY}
We have performed a comprehensive study of the Floquet-Hofstadter butterfly spectrum in (1+3+1)-chiral twisted multilayer graphene (CTMLG) under the influence of a uniform perpendicular magnetic field. Our study explored both the equilibrium case under the effects of different forms of light, providing insights into how external conditions and structural configurations affect the electronic properties of this intriguing material. We found significant discrepancies between the effects of symmetric ($\theta_1 = \theta_2$) and asymmetric ($\theta_1 \neq \theta_2$) twist angle configurations on the electronic structure. 
{These findings allow a deeper understanding of the role of symmetry in layered materials.}
In equilibrium, the symmetric configuration maintained a clear fractal Hofstadter butterfly spectrum, showing a uniform interlayer coupling. This uniformity results in a consistent and predictable electronic response, typical for systems with balanced interactions. In contrast, the asymmetric configuration caused noticeable changes in the electronic spectrum due to uneven interlayer interactions. These changes highlight the system's sensitivity to structural irregularities, which disrupt uniformity and create more complex spectral features. This difference between symmetric and asymmetric configurations emphasizes the importance of controlling the twist angle when designing materials with specific electronic properties.

When we introduced circularly polarized light into the system, we observed significant changes in the Hofstadter butterfly spectrum as the driving strength increased. The chiral symmetry-breaking effects of the light were particularly pronounced for asymmetric twist configurations, where the spectrum became increasingly asymmetric around $E = 0$, especially at the central $n=0$ Landau level. 
This asymmetry shows how external perturbations can enhance the inherent differences in electronic behavior between symmetric and asymmetric systems. 
It also highlights the potential usage of light as a tool for manipulating quantum states and breaking symmetries in a controlled manner.
In contrast, it was found that the linearly polarized waveguide light preserved the chiral symmetries of the system for both symmetric and asymmetric twist angle configurations.
However, the effect on the bandwidth was far from straightforward. At lower driving strengths, the bandwidth decreases, suggesting suppression of electronic interactions. At higher driving strengths, the bandwidth increases, indicating dynamic modulation of the system by the external field. This non-monotonic behavior reflects the complex interplay between light and matter, where the response of the system depends critically on the intensity and nature of the external stimulus.

These results deepen our understanding of the Floquet-Hofstadter spectrum in twisted graphene systems. They also open new avenues for exploring how symmetry, interlayer coupling, and external potentials combine to produce new effects.
Our work shows how light and structural changes can tune the electronic properties. This provides a basis for designing advanced materials with specific quantum behaviors. 
This research combines fundamental physics with practical applications, providing insights that can inspire future developments in quantum electronics and photonics.

\section*{Acknowledgment}
P.D. and D.L. acknowledge partial financial support from FONDECYT 1231020.

\appendix

\section{Incorporating magnetic field into the (1+3+1)-CTMLG Hamiltonian}\label{appendix A}
We introduce a perpendicular magnetic field $\boldsymbol{B} = B_z \boldsymbol{e}_z$ into the Hamiltonian describing CTMLG using the minimal substitution $\boldsymbol{\hat{p}} \longrightarrow \boldsymbol{\hat{p}} + e\boldsymbol{A}.$ For our purposes, it is suitable to use the Landau gauge $\boldsymbol{A} = B(-y, 0).$ In this setup, we find that the intra-layer blocks of the Hamiltonian can be expressed as
\begin{equation}
	H\left(\frac{\theta}{2}\right)=\omega_{c}\left(\sigma^{+}e^{i\frac{\theta}{2}}a+\sigma^{-} e^{-i\frac{\theta}{2}}a^{\dagger}\right)
\end{equation}
and the ladder operators are given by
\begin{align}
&	a=\frac{\ell}{\sqrt{2}}\left(p_x-eBy-ip_y  \right)\\
&	a^\dagger=\frac{\ell}{\sqrt{2}}\left(p_x-eBy+ip_y  \right)
\end{align}
fulfilling the  commutation relation $\left[a, a^{\dagger}\right]=1$, 
where the $ \omega_c=\sqrt{2}v_F/\ell$ is the cyclotron energy, and  $ \ell=1/\sqrt{eB}$ is the magnetic length.

It is now convenient to express the full Hamiltonian in a basis of layer index  $L=1, 2, \cdots 5$, sublattice {$\sigma= A, B$}, guiding center $y$ and Landau level $n$ degrees of freedom using a vector  $\ket{ L,n,\sigma, y}$.  The intralayer Hamiltonian then is given by
\begin{equation}
	\eqfitpage{H\left(\frac{\theta}{2}\right)=\omega_{c} \sum\limits_{L, n, y}\left(e^{-i\theta/2} \sqrt{n+1}\ket{L, n+1, A, y}\bra{L,n, B, y}\right)+\mathrm{hc}}.
\end{equation}
The interlayer Hamiltonian in the same basis can be expressed as
\begin{widetext} 
	\begin{align}
	T_{12}(\mathbf{k})=&\sum_{n',n, \sigma, \sigma', j}\left[T_{1} F_{n' n}\left(\mathbf{z}_1\right) e^{-i k_x ,k_{\theta_1} \ell^2} e^{-4 \pi i \frac{p}{q} j}\ket{2, n', \sigma, j}\bra{ 1, n ,\sigma', j}\right.\\
    &+T_{2}F_{n' n}\left(\mathbf{z}_{2}\right) e^{i k_{y} \Delta} e^{\frac{i}{2} k_x k_{\theta_1} \ell^2} e^{i \pi \frac{p}{q}(2 j-1)}\ket{ 2 ,n', \sigma, j+1}\bra{ 1, n, \sigma', j} \nonumber \\
	&\left.+T_{3}F_{n' n'}\left(\mathbf{z}_{3}\right) e^{-i k_{y} \Delta} e^{\frac{i}{2} k_x k_{\theta_1} \ell^2} e^{i \pi \frac{p}{q}(2 j+1)}\ket{ 2, n', \sigma, j-1}\bra{ 1, n, \sigma', j}\right],\nonumber \\
    T_{45}(\mathbf{k})=&\sum_{n',n, \sigma, \sigma', j}\left[T_{1} F_{n' n}\left(\mathbf{z}_1\right) e^{-i k_x ,k_{\theta_2} \ell^2} e^{-4 \pi i \frac{p}{q} j}\ket{5, n', \sigma, j}\bra{ 4, n ,\sigma', j}\right.\\
    &+T_{2}F_{n' n}\left(\mathbf{z}_{2}\right) e^{i k_{y} \Delta} e^{\frac{i}{2} k_x k_{\theta_1} \ell^2} e^{i \pi \frac{p}{q}(2 j-1)}\ket{ 5 ,n', \sigma, j+1}\bra{ 4, n, \sigma', j} \nonumber \\
	&\left.+T_{3}F_{n' n'}\left(\mathbf{z}_{3}\right) e^{-i k_{y} \Delta} e^{\frac{i}{2} k_x k_{\theta_2} \ell^2} e^{i \pi \frac{p}{q}(2 j+1)}\ket{ 5, n', \sigma, j-1}\bra{ 4, n, \sigma', j}\right],\nonumber 
\end{align}	
\end{widetext}
	where $p/q$ is a rational number, with $p, q \in  \mathbb{Z}$, namely
	\begin{equation}
		\Phi=\frac{p}{q} \Phi_{0}, \quad \phi_{0}=\frac{h c}{e}.
	\end{equation}
	The integers $p$ and $q$ have a significant meaning  within the magnetic Brillouin zone (MBZ), indeed
	\begin{equation}
		\frac{\Phi}{\Phi_0}=\frac{B\Omega_M}{\Phi_0}=\frac{p}{q}=\alpha
	\end{equation}
	where $\Omega_M=16\pi^2/\sqrt{3}k_{\theta}^2$ is the moiré unit cell area, the flux through a moiré unit cell increases 
	as $\theta$  decreases.  It turns out that for a given $\alpha$, the spectrum
	consists of a group of bands with gaps, and for $\alpha\precsim1$, i.e. larger fields, the pattern of gaps shows a meaningful splitting between Landau levels (Fig. \ref{equilibrium-case1}). 
and $\Delta=\sqrt{3}k_\theta \ell^2/2$ is  the change in the guiding center in a tunneling process, in this case, the guiding center rewritten as  $y = y_0 + (mq + j) \Delta$,  $m$ is a parameter in Fourier transform, with
\begin{equation}
	0<y_{0}=k_{x} \ell^{2}<\Delta, \quad 0<j<q-1
\end{equation}
and $j=j+q$. The momentum $\mathbf{k} = (k_x, k_y)$ lies inside the “magnetic” Brillouin zone 
\begin{equation}
	0<k_{x} k_{\theta} \ell^{2}=k_{\theta} y_{0}<\frac{4 \pi p}{q}, \quad 0< k_y<\frac{4 \pi}{\sqrt{3}k_{\theta}q}
\end{equation}
and the function $F$  is
\begin{equation}
	\eqfitpage{F_{n' n}(\mathbf{z})=
		\begin{cases}
			\sqrt{\frac{n !}{n' !}}\left(-z_{x}+i z_{y}\right)^{n'-n} e^{-\frac{z^{2}}{2}} \mathcal{L}_{n}^{n'-n}\left(z^{2}\right) & n'
			\geq n \\
			\sqrt{\frac{n' !}{n !}}\left(z_{x}+i z_{y}\right)^{n-n'} e^{-\frac{z^{2}}{2}} \mathcal{L}_{n'}^{n-n'}\left(z^{2}\right) & 
			n' < n		
	\end{cases}}
\end{equation}
where  $ \mathbf{z}_{j}=\frac{q_{j x}+i q_{j y}}{\sqrt{2}}\ell$, and $\mathcal{L}$ being the associated Laguerre polynomial.
\bibliographystyle{unsrt}
\bibliography{mybib}
\appendix
\end{document}